\documentclass{article}
\usepackage{graphicx}
\usepackage[utf8]{inputenc}
\usepackage{eucal}
\usepackage{amsmath,amssymb}
\usepackage{hyperref}
\usepackage{epstopdf}
\usepackage{slashed}
\usepackage{ulem}

\usepackage{xcolor}
\definecolor{orange}{RGB}{255,127,0}
\definecolor{brown}{RGB}{102,51,0}
\definecolor{myred}{RGB}{192,0,0}
\definecolor{Darkgreen}{RGB}{30,120,30}
\definecolor{Darkblue}{RGB}{0,0,200}
\newcommand{\comment}[1]{}

\newcommand\lsim{\mathrel{\rlap{\lower4pt\hbox{\hskip1pt$\sim$}}
    \raise1pt\hbox{$<$}}}
\newcommand\gsim{\mathrel{\rlap{\lower4pt\hbox{\hskip1pt$\sim$}}
    \raise1pt\hbox{$>$}}}
\newcommand{\ba}{\begin{array}}
\newcommand{\ea}{\end{array}}

\newcommand{\be}{\begin{equation}}
\newcommand{\ee}{\end{equation}}
\newcommand{\bear}{\begin{eqnarray}}
\newcommand{\eear}{\end{eqnarray}}

\newcommand{\ket}{\,\rangle}
\newcommand{\bra}{\langle \,}

\newcommand{\mB}{\mathcal{B}}

\newcommand{\mF}{\mathcal{F}}
\newcommand{\mG}{\mathcal{G}}

\newcommand{\mM}{\mathcal{M}}

\newcommand{\mO}{\mathcal{O}}

\newcommand{\invmass}{m_{\pi^- e^+e^-}}

\title{Improved description of di-lepton production in $\tau^- \to \nu_{\tau}P^-$ decays
}
\author{Adolfo Guevara$^1$ $^2$, Gabriel L\'opez Castro$^3$ and Pablo Roig$^3$\\
$^1$ Departamento de F\'isica At\'omica, Molecular y Nuclear\\
and Instituto Carlos I de F\'isica Te\'orica y Computacional\\
Universidad de Granada, E-18071 Granada, Spain.\\
$^2$ Departament de F\'isica Te\`orica, IFIC,\\ Universitat de Val\`encia - CSIC,\\
Apt. Correus 22085, E-46071 Val\`encia, Spain\\
$^3$ Departamento de F\'\i sica, Centro de Investigaci\'on \\ y de Estudios Avanzados del 
Instituto Polit\'ecnico Nacional. \\ AP 14-740, 07000, Ciudad de M\'exico, M\'exico}

\graphicspath{{Figs/}}

\begin{document}

\maketitle

{\bf Abstract} Recently, the Belle collaboration reported the first measurements of the $\tau^- \to\nu_{\tau}\pi^-e^+e^-$ branching fraction and the spectrum of the pion-dielectron system. {In an analysis previous to Belle's results, we evaluated this branching fraction which turned out to be compatible with that reported by Belle, although with a large uncertainty.} This is the motivation to seek for improvements on our previous evaluation of $\tau^- \to \nu_{\tau}\pi^-\ell^+\ell^-$ decays ($\ell=e,\,\mu$). In this paper we improve our calculation of the $WP^-\gamma^*$ vertex by including flavor symmetry breaking effects in the framework of the Resonance Chiral Theory. We impose QCD short-distance behavior to constrain most parameters and data on the $\pi^-e^+e^-$ spectrum reported by Belle to fix the remaining free ones. As a result, improved predictions for the branching ratios and hadronic/leptonic spectra are reported, in good agreement with observations. Analogous calculations for the strangeness-changing $\tau^- \to \nu_{\tau}K^-\ell^+\ell^-$ transitions are reported for the first time. {Albeit one expects the $m_{\pi\mu^+\mu^-}$ spectrum to be measured in Belle-II and the observables with $\ell=e$ can be improved, it is rather unlikely that the $K$ channels can be measured due to the suppression factor $|V_{ud}/V_{us}|^2=0.05$.}

\section{Introduction}\label{sec:introduction}

The search for signals of physics beyond the Standard Model (SM) 
requires a good understanding of SM processes either to discard possible backgrounds coming from it, such as large radiative corrections \cite{Guevara:2016trs, Husek:2017vmo,  Kampf:2018wau} or 
to have under good control 
hadronic contamination in precision tests of the SM \cite{Guevara:2015pza}. 
In addition to offering a clean laboratory to test the hadronization of the weak currents, {some} semileptonic $\tau$ lepton decays{, such as $\tau\to\nu_\tau P (\gamma)$ for $P=\pi,K$,} provide a good example where SM effects can be reliably calculated to disentangle possible New Physics signals hidden in precision observables. \\

 In  Ref. \cite{Guevara:2013wwa} we reported 
 the first prediction of  $\mB(\tau\to\nu_\tau\pi\ell\overline{\ell})$ and the corresponding di-lepton spectrum where $\ell=e,\mu$ (
 this can be viewed as the crossed channels of lepton pairs produced in $\pi_{\ell2}$ decays \cite{Bijnens:1992en} in a larger kinematical domain
 ); 
 later on the Belle collaboration \cite{Jin:2019goo} announced the first searches of these decays. Recently, some of us have also reported similar studies 
 of $\tau^-\to\nu_\tau\pi^-\pi^0\ell\bar{\ell}$ decays \cite{GutierrezSantiago:2020bhy}. 
 Together with the five lepton decays of tau leptons \cite{5leptons}, they provide a better description of possible backgrounds in Lepton Number- or Lepton Flavor Violation searches in $\tau$ decays
 . 
 Motivated by the Belle Collaboration studies \cite{Jin:2019goo}, in this work we revisit our predictions for $\tau\to\nu_\tau\pi\ell\overline{\ell}$ decays with the aim of improving the theoretical description of structure-dependent effects and to get 
 reduced 
 uncertainties. In addition, we make  
 for the first time 
 an analogous analysis of the strangeness-changing processes $\tau\to\nu_\tau K\ell\overline{\ell}$ as well.\\

In these phenomena, the $W\gamma^\star P$ vertex plays a central role and its description is necessary to understand the radiative corrections to the $\tau^-\to\nu_\tau P^-$ decays \cite{Arroyo-Urena:2021nil}. 
This vertex also involves 
parameters which are needed to describe the pion transition form factor (TFF),  
which is required to compute the dominant piece (the pion pole) of the hadronic light-by-light contribution to 
the anomalous magnetic moment of the $\mu$ lepton, $a_\mu$. {(The TFF can be obtained by our vector form factor (see section \ref{sec:Form_Factors}) by considering Bos\'e symmetry.) Although knowledge on these parameters could in principle help reduce the uncertainty on the hadronic part of $a_\mu$ \cite{Aoyama:2020ynm}, the $\tau^-\to\nu_\tau\pi^-e^+e^-$ data does not (and is not foreseeable to) have the necessary precision to improve actual predictions on the $\pi$-pole contribution to $a_\mu$.}\\

The problem with the description of these effective vertices arises when one tries to describe them in terms of the fundamental fields of the Standard Model, since at energies below the $m_\tau$ scale, one can not give a proper perturbative description of color interactions. 
{However, the decay amplitude involving these vertices can be, for the sake of convenience, split into a part where the hadronic current $\langle0|\overline{u}\gamma_\mu(1-\gamma_5)d|\pi^-\rangle=-i\sqrt{2}f_\pi p_\mu$ and the electromagnetic interactions are computed using scalar QED (sQED), which we call structure-independent, and a part where more involved hadronic interactions are computed using an Effective Field Theory, called structure-dependent.}
Thus, we try to surpass the difficulties of calculating the structure-dependent part using Resonance Chiral Theory (R$\chi$T) \cite{Ecker:1988te,Ecker:1989yg}, which is an extension of Chiral Perturbation Theory ($\chi$PT) \cite{Gasser:1983yg,Gasser:1984gg,Weinberg:1978kz} that includes resonances as active degrees of freedom. $\chi$PT relies on the chiral symmetry group $G=U(3)_L\otimes U(3)_R$ of the massless QCD Lagrangian. After it gets spontaneously broken, $G\to U(3)_V$, the remaining symmetry gets explicitly broken when the masses of the light quarks are considered to be non-vanishing. The $\mB(\tau^-\to\nu_\tau\pi^-\ell\overline{\ell})$ and di-lepton spectrum were computed previously in ref. \cite{Guevara:2013wwa} using such techniques, however, the novelty in 
the present 
treatment is that we include the effects of finite different light-quark masses as done for the Transition Form Factor of the pseudo Goldstone bosons for the Hadronic Light-by-Light part of the $a_\mu$ in ref. \cite{Guevara:2018rhj} (over \cite{Roig:2014uja}, where these were neglected). We also give a more thorough treatment of the uncertainties than those in ref. \cite{Guevara:2013wwa}, thus obtaining consistent results comparing with the corresponding form factors given in ref. \cite{Cirigliano:2006hb}. Furthermore, the recent measurement of the branching fraction with a lower limit in the invariant mass of the pion and di-lepton pair \cite{Jin:2019goo}, $m_{\pi e^-e^+}$, motivates further this re-analysis, since in the $m_{\pi e^-e^+}\geq1.05$ GeV region the branching fraction gets saturated by the structure-dependent contribution. 
While most 
of the parameters of the model can be constrained by means of the high-energy behavior of QCD, some of them remain loose.
 We fit these to the measured  invariant mass $m_{\pi^- e^-e^+}$ spectra and also to the measurement of the branching fraction $\mB(m_{\pi^- e^-e^+}\geq1.05 \text{ GeV})=(5.90\pm1.01)\times10^{-6}$ \cite{Jin:2019goo}. {Despite the access to the invariant mass spectra data for the $\tau^+\to\nu_\tau\pi^+\overline{\ell}\ell$ decay, we will only make use of the data for the $\tau^-$ decay. The reason not to use both sets is that the spectra have incompatibilities in several bins; also, when fitting individually the $\pi^+$ data set leads to unphysical conditions (see discussion in subsection \ref{sec:Fit}).} As a result, we improve our predictions, with correspondingly reduced uncertainties.\\
 
 The outline of the paper 
 is as 
 follows. In section \ref{sec:amplitudes} the different contributions to the matrix element are collected. In section \ref{sec:Structure_dependent} we introduce the Lagrangian used for computing the structure-dependent corrections, calculate the corresponding form factors (including flavor-breaking corrections to our previous results) and derive the short-distance constraints among resonance couplings. In section \ref{sec:Pheno_anal} we carry out our phenomenological analysis, including a fit to Belle $\tau^-\to\pi^- e^+ e^- \nu_\tau$ data and predicting the partner $(\pi\leftrightarrow K, e\leftrightarrow \mu)$ modes, yet to be discovered. We 
 give our conclusions 
 in section \ref{sec:concl}.

 \section{Amplitudes}\label{sec:amplitudes}
 
\begin{figure}[!ht]
 \centering\includegraphics[scale=0.24]{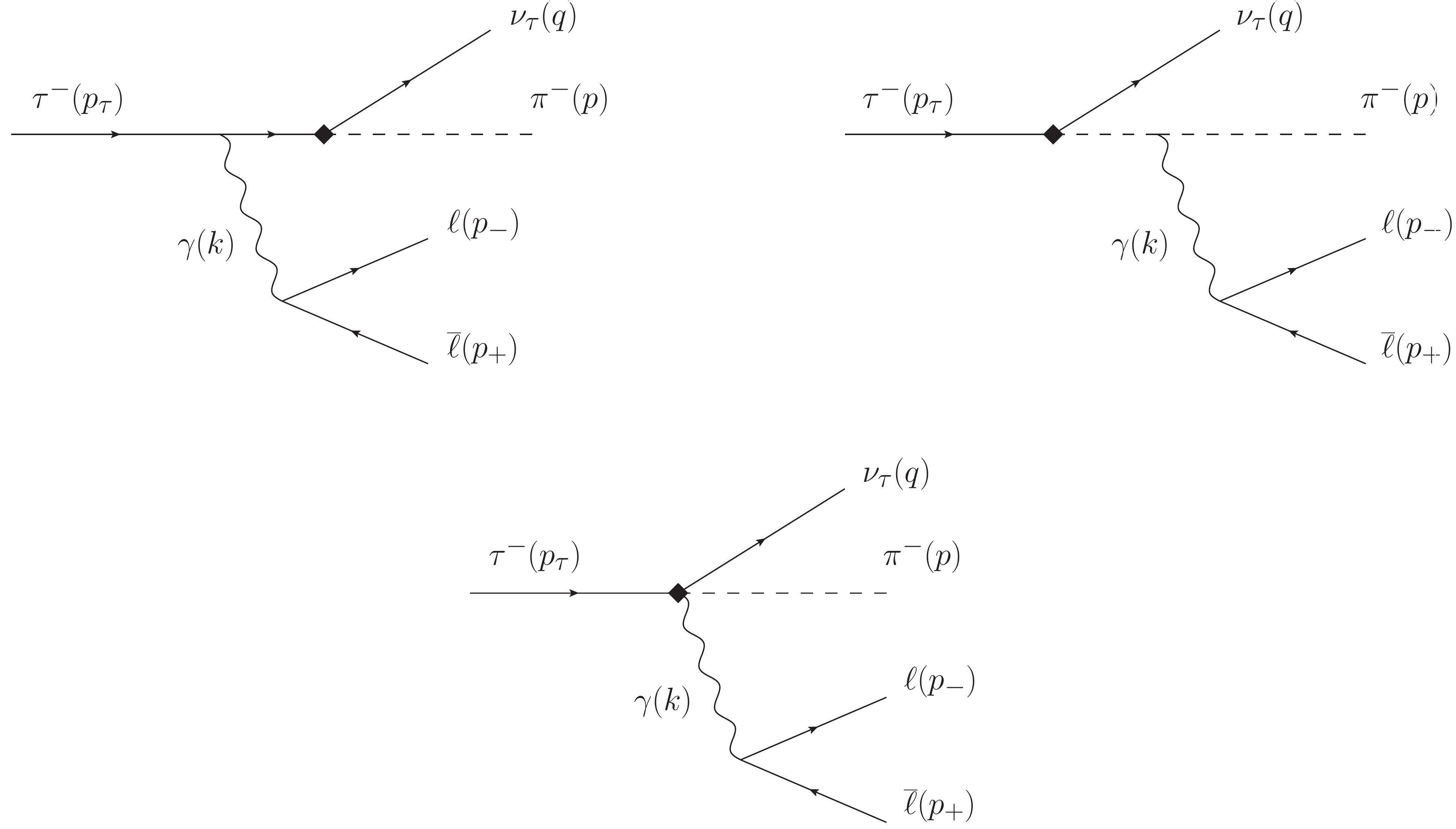}\caption{Feynman diagrams of the SI contributions (only scalar QED is used for the radiation off the $P^-$ meson) to the $\tau^-(p_\tau)\to\nu_\tau(q)P^-(p)\ell(p_-)\overline{\ell}(p_+)$ decay amplitude. The diamond vertex is an effective vertex meaning the $W$ boson has been integrated out.}\label{fig:IB_diagrams}
 \end{figure}

 For convenience, we take three kinds of contributions to the decay amplitude: the first called inner bremsstrahlung (IB) or structure independent (SI). The other two are the structure dependent (SD) ones, namely the polar- (V) and axial-vector (A) parts of the left-handed weak charged current. The IB amplitude {can be obtained using the sQED Lagrangian,} 
where the photon is either radiated by the $\tau$ lepton, off the pseudo Goldstone boson ($\pi$ or $K$) or by the longitudinal propagation mode of the $W^-$ boson, a contribution which is needed to achieve gauge invariance of the total IB amplitude. The total IB contribution is shown in eq. (\ref{eq:amplitudes}), along with the parametrization of the SD parts as given in ref. \cite{Guevara:2013wwa}. The momenta definition is given in Figure \ref{fig:IB_diagrams}.
 
 The different contributions to the matrix element are ($D=d,s$ for $P=\pi,\,K$)
 
 \begin{subequations}\label{eq:amplitudes}
 \begin{align}
     \mM_{IB}&=-iG_FV_{uD}f_\pi m_\tau\frac{e^2}{k^2}J_\ell^\nu  \overline{u}_{\nu_\tau}(1+\gamma_5)\left[\frac{2p_\nu}{2 p\cdot k+k^2}+\frac{2{p_\tau}_\nu-\slashed{k}\gamma_\nu}{-2p_\tau\cdot k+k^2}\right]u_\tau,\\
     \mM_{V}&=-G_FV_{uD}\frac{e^2}{k^2}J_\ell^\nu J_\tau^\mu F_V(W^2,k^2)\varepsilon_{\mu\nu\alpha\beta}k^\alpha p^\beta,\\
     \mM_{A}&=iG_FV_{uD}\frac{e^2}{k^2}J_\ell^\nu J_\tau^\mu\left\{F_A(W^2,k^2)\left[(W^2+k^2-m_\pi^2)g_{\mu\nu}-2k_\mu p_\nu\right]\frac{}{}\nonumber\right.\\
     &\left.\hspace*{18ex}-\frac{}{}A_2(W^2,k^2)k^2g_{\mu\nu}+A_4(W^2,k^2)k^2(p+k)_\mu p_\nu\right\}.\label{eq:amplitudes_MA}
 \end{align}
 \end{subequations}
 
 Here, $J_\ell^\nu=\overline{u}(p_-)\gamma^\nu v(p_+)$ and $J_\tau^\mu=\overline{u}(q)(1+\gamma_5)\gamma^\mu u(p_\tau)$ are the lepton electromagnetic and $\tau$ weak charged currents, respectively. We use $W^2\equiv(p_\tau-q)^2$ and $k^2\equiv(p_-+p_+)^2$ as the two independent Lorentz-invariants upon which the form factors ($F_V, F_A, A_2, A_4$) depend. In ref. \cite{Guevara:2013wwa} the axial amplitude was given only in terms of three form factors ($F_V,F_A$ and a combination of $A_2$ and $A_4$ called $B$), since at chiral order $p^4$, the $A_2$ and $A_4$ form factors are linearly dependent and can be written in terms of  the pseudo Goldstone electromagnetic form factor $F_V^P(k^2)$ \cite{Bijnens:1992en}. Here, $A_2$ and $A_4$ cannot be recast in terms of $F_V^P(k^2)$, since we are considering contributions of chiral order $p^6$. Furthermore, including the complete set of leading-order chiral symmetry breaking contributions will change the pion pole for the massive pion propagator. As a result, the $A_2$ and $A_4$ form factors become linearly independent and the axial-vector part of the left hadronic current cannot be expressed in terms of the two form factors $\mF(W^2,k^2,p^2)$ and $\mG(W^2,k^2,p^2)$ of refs. \cite{Cirigliano:2006hb,Knecht:2001xc,Cirigliano:2004ue} (see discussion after eq. (\ref{eq22})). 
 
 \section{Structure dependent form factors}\label{sec:Structure_dependent}
 
 \subsection{The relevant operators}\label{sec:Relavant_operators}
 
 In this section we will present, for the sake of simplicity, only the relevant operators in the R$\chi$T Lagrangian needed to compute the form factors, which are given in the next subsection. We will be concise here, for a more extended discussion see eg. ref. \cite{Guevara:2018rhj}. R$\chi$T extends the domain of applicability of Chiral Perturbation Theory \cite{ Weinberg:1978kz, Gasser:1983yg, Gasser:1984gg} ($\chi$PT) by adding the light-flavored resonances as active degrees of freedom.\\ 

We start with operators involving no resonances, these being~\footnote{Although these terms also appear in the $\chi$PT Lagrangian, their couplings get shifted in the presence of resonance contributions (see for instance \cite{Bernard:1991zc,SanzCillero:2004sk,Guo:2014yva}).}

\begin{equation}
    \mathcal{L}_{0\,Res}=\frac{f^2}{4}\langle u^\mu u_\mu + \chi_+\rangle+\mathcal{L}_{WZW}+C_7^W\mathcal{O}_7^W+C_{11}^W\mathcal{O}_{11}^W+C_{22}^W\mathcal{O}_{22}^W\,,
\end{equation}
 where the first term is given by the leading $\chi$PT Lagrangian operators of chiral order $p^2$ \cite{ Weinberg:1978kz, Gasser:1983yg, Gasser:1984gg}, the second one is the anomalous Wess-Zumino-Witten Lagrangian of $\mathcal{O}(p^4)$ \cite{Wess:1971yu, Witten:1983tw} and the last three operators belong to the subleading odd-intrinsic parity sector $\mathcal{O}(p^6)$ Lagrangian \cite{Bijnens:2001bb}. We neglect operators not included in this Lagrangian.  Congruently with refs.  \cite{Cirigliano:2006hb}, \cite{Cirigliano:2004ue} and \cite{Kampf:2011ty}, we will not consider any $\mathcal{O}(p^8)$ contribution whatsoever. In the first term, $f$ is the decay constant in the chiral limit, which we will set to $f=f_\pi\sim92$ MeV, $u^\mu$  and $\chi_+$ are chiral tensors \cite{Bijnens:1999sh}, the former containing derivatives of the $\pi/K$ fields and external spin-one currents and the latter scalar currents involving the previous fields masses squared, $m_{\pi/K}^2$, times even powers of such fields. \\

 The equations of motion of the resonances give their classical fields in terms of series of chiral tensors of different order. The resonances are said to be integrated out {(tree-level integration)} when the classic fields are substituted in favor of chiral tensors in the resonant Lagrangian. Integrating the resonances out using the leading-order terms of the equations of motion {very approximately} saturates the $\mathcal{O}(p^4)$ (and leading  $\mathcal{O}(p^6)$) contributions in the even-intrinsic parity sector \cite{Ecker:1988te, Ecker:1989yg, Cirigliano:2006hb}; therefore, we will not use the non-resonant $\mathcal{O}(p^4)$ set of operators {for the sake of simplicity, since they are considered to yield negligible contributions.}
 Since we will only consider leading-order terms in the resonances equations of motion, the $\mathcal{O}(p^6)$ chiral low-energy constants in the odd-intrinsic parity sector cannot be saturated upon resonance exchange \cite{Kampf:2011ty}, therefore we have to include the three contributing $C_i^W \mathcal{O}_i^W$ terms \cite{Bijnens:2001bb}:
 \begin{eqnarray}
 \mathcal{O}_7^W & = & i \epsilon_{\mu\nu\alpha\beta} \bra \chi_- f_+^{\mu\nu} f_+^{\alpha\beta}\ket,
    \nonumber\\
    \mathcal{O}_{11}^W & =&  i \epsilon_{\mu\nu\alpha\beta} \bra \chi_+ [f_+^{\mu\nu}, f_-^{\alpha\beta}]\ket,
    \\
    \mathcal{O}_{22}^W & = & i \epsilon_{\mu\nu\alpha\beta} \bra u^\mu \{ \nabla_\rho f_+^{\rho\nu}, f_+^{\alpha\beta}\} \ket\,,\nonumber
 \end{eqnarray}
 where the following chiral tensors \cite{Bijnens:1999sh} enter: $\chi_-$ gives odd powers of the $\pi/K$ fields with factors involving $m_{\pi}^2$ or $m_K^2$, $\nabla_\mu$ is the covariant derivative and includes spin-one left and vector external currents through the connection and $f_{\pm}^{\mu\nu}$ yields the field-strength tensors of the charged-weak or electromagnetic fields.\\
 
 We turn next to those operators with one resonance field, in either intrinsic parity sector,
 \begin{equation}
\mathcal{L}_{1\,Res}= \mathcal{L}_{1\,Res}^{even}+\mathcal{L}_{1\,Res}^{odd}\,.    
 \end{equation}
 In turn, the first piece can be further divided according to the quantum numbers of this resonance 
 \begin{equation}
 \mathcal{L}_{1\,Res}^{even} = \sum_{R_i=V,A,P}\mathcal{L}_{1\,R_i}^{even}\,.
 \end{equation}
 The contributions with one vector resonance read~\footnote{$V_{\mu\nu}$ (analogously $A_{\mu\nu}$ for axial resonances below) is a matrix in flavor ($u,d,s$) space and we use the antisymmetric tensor formalism for spin-one fields for convenience \cite{Ecker:1988te,Ecker:1989yg}.} \cite{Ecker:1988te, Cirigliano:2006hb}
 \begin{equation}\label{eq:1R_V_even}
 \mathcal{L}_{1\,V}^{even}=\frac{F_V}{2\sqrt{2}}\bra V_{\mu\nu} f_+^{\mu\nu}\ket +\frac{i}{2\sqrt{2}} G_V\bra V_{\mu\nu} [u^\mu,u^\nu]\ket+ \frac{\lambda_V}{\sqrt{2}}\bra V_{\mu\nu} \{  f_+^{\mu\nu}, \chi_+\}\ket \,,
 \end{equation}
   {where the $V$ field (we assume ideal mixing of neutral mesons) has an analogous flavor structure as the pseudo Goldstone field $\phi$, namely
   \begin{equation}
    V_{\mu\nu}=\left(\begin{array}{ccc}\frac{1}{\sqrt{2}}\left(\rho^0_{\mu\nu}+\omega_{\mu\nu}\right)&\rho^+_{\mu\nu}&{K^\star}^+_{\mu\nu}\\\rho^-_{\mu\nu}&\frac{1}{\sqrt{2}}\left(-\rho^0_{\mu\nu}+\omega_{\mu\nu}\right)&{K^\star}^0_{\mu\nu}\\{K^\star}^-_{\mu\nu}&{\overline K^{\hspace*{.5ex}\star}}^0_{\mu\nu}&\phi_{\mu\nu}\end{array}\right).
   \end{equation}
   In eq. (\ref{eq:1R_V_even}),} the first two operators {give the contribution from the coupling of vector resonances to external fields in the chiral limit and the last term gives the flavor-breaking corrections to such couplings. }
   Our $\lambda_V=\sqrt{2}\lambda_6^V$, using the notation in ref. \cite{Cirigliano:2006hb}. This last operator is the only one included from the full basis of $\mathcal{O}(p^4)$ even-intrinsic parity operators in ref. \cite{Cirigliano:2006hb} since it is the single one that can contribute to the $U(3)_V$ breaking in the $V-\gamma$ coupling. There are, however, two reason to disregard basis of operators in the even-intrinsic parity sector: The operators that are relevant to the process can be dismissed on the basis of resonance field redefinitions\footnote{
   {Through the redefinition of the vector resonance field $V\to V+g\{V,\chi_+\}$ it is possible to cancel the $\lambda_V$ operator \cite{Cirigliano:2006hb}},
   however, we keep it in order to show the full basis of possible $U(3)_V$ breaking operators since we do not consider the full even-intrinsic parity basis of ref. \cite{Cirigliano:2006hb}. We will show later that this is consistent, since the short distance constraints give $\lambda_V=0$.}; if we, however, keep such operators, they will only give subleading contributions to those from the first two operators in eq. (\ref{eq:1R_V_even}) with no contribution to $U(3)_V$-breaking vertices.\\
   
   The axial resonance operators present a similar feature {and an analogous flavor-space structure to that of the vector mesons}. This is, the $\mathcal{O}(p^4)$ one-resonance even-intrinsic parity operators for axial resonances in ref. \cite{Cirigliano:2006hb} can be absorbed through field redefinitions. We will therefore disregard any contribution from this part of the Lagrangian, including the $U(3)_V$ breaking terms to the axial-vector resonance coupling to external currents, namely, the $JA$ vertex.
 The remaining contributions with one resonance field are \cite{Ecker:1988te}
 \begin{equation}\label{eq:1R_A_even}
 \mathcal{L}_{1\,A/P}^{even}=\frac{F_A}{2\sqrt{2}}\langle A_{\mu\nu} f_-^{\mu\nu}\rangle +id_m\langle P \chi_- \rangle,
 \end{equation}
 with $P$ a matrix in three-flavors space containing the lightest pseudoscalar resonances. The inclusion of the pseudoscalar resonance is necessary in order to obtain consistent short-distance constraints in $\langle VAP\rangle$ and $\langle VJP\rangle$ Green's functions \cite{Cirigliano:2006hb,Kampf:2011ty,Roig:2013baa,Kadavy:2020hox}.  All Feynman diagrams involving these resonances will give $U(3)$ breaking contributions to the amplitude due to the last term in eq. (\ref{eq:1R_A_even}). We have neglected other spin-zero resonance contributions (scalar and heavier pseudoscalar resonances \cite{Guevara:2013wwa}), which are not needed for theoretical consistency and are irrelevant phenomenologically. 
 The odd-intrinsic parity contributions to $\mathcal{L}_{1\,Res}$ are \cite{RuizFemenia:2003hm}~\footnote{Since we are only  considering  operators with one $\pi/K$ field, these constitute a basis. In the general case, the basis is given in ref. \cite{Kampf:2011ty}. The translation between them can be read from ref. \cite{Roig:2013baa}.}
 \begin{equation}
\mathcal{L}_{1\,Res}^{odd}=\sum_{j=1}^7 \frac{c_j}{M_V}\mO^j_{V}+\varepsilon_{\mu\nu\alpha\beta}\langle \kappa_5^P\{f_+^{\mu\nu}, f_+^{\alpha\beta}\}P\rangle\,,
 \end{equation}
 with the operators
 \begin{eqnarray}
   \mathcal{O}^1_{V}&=&\varepsilon_{\mu\nu\rho\sigma}\langle\left\{V^{\mu\nu},f^{\rho\alpha}_+\right\}\nabla_\alpha u^\sigma\rangle\,,\nonumber\\
  \mathcal{O}^2_{V}&=&\varepsilon_{\mu\nu\rho\sigma}\langle\left\{V^{\mu\alpha},f^{\rho\sigma}_+\right\}\nabla_\alpha u^\nu\rangle\,,\nonumber\\
  \mathcal{O}^3_{V}&=&i\varepsilon_{\mu\nu\rho\sigma}\langle\left\{V^{\mu\nu},f^{\rho\sigma}_+\right\}\chi_-\rangle\,,\\
  \mathcal{O}^4_{V}&=&i\varepsilon_{\mu\nu\rho\sigma}\langle V^{\mu\nu}\left[f^{\rho\sigma}_-,\chi_+\right]\rangle\,,\nonumber\\
  \mathcal{O}^5_{V}&=&\varepsilon_{\mu\nu\rho\sigma}\langle\left\{\nabla_\alpha V^{\mu\nu},f^{\rho\alpha}_+\right\}u^\sigma\rangle\,,\nonumber\\
  \mathcal{O}^6_{V}&=&\varepsilon_{\mu\nu\rho\sigma}\langle\left\{\nabla_\alpha V^{\mu\alpha},f^{\rho\sigma}_+\right\}u^\nu\rangle\,,\nonumber\\
  \mathcal{O}^7_{V}&=&\varepsilon_{\mu\nu\rho\sigma}\langle\left\{\nabla^\sigma V^{\mu\nu},f^{\rho\alpha}_+\right\}u_\alpha\rangle.\nonumber
 \end{eqnarray}

 In the following, we quote those terms bilinear in resonance fields (we do not display the kinetic terms for the resonances, which can be found in ref. \cite{Ecker:1988te}, as they do not contribute to the effective vertices).
 
 \begin{equation}
     \mathcal{L}_{2\,Res}=\mathcal{L}_{2\,Res}^{even}+\mathcal{L}_{2\,Res}^{odd}\,,
 \end{equation}
 with \cite{Cirigliano:2006hb,GomezDumm:2003ku,Cirigliano:2003yq,Guo:2009hi}~\footnote{The operator with coefficient $e^V_M$ allows to account for $U(3)$ breaking effects in the vector resonance masses, in agreement with phenomenology.}
 \begin{equation}
\mathcal{L}_{2\,Res}^{even}=-e^V_M\langle V_{\mu\nu}V^{\mu\nu}\chi_+\rangle+\lambda_1^{PV}\mathcal{O}_1^{PV}+\lambda_2^{PV}\mathcal{O}_2^{PV}+\lambda_1^{PA}\mathcal{O}_1^{PA}+\sum_{i=1}^5\lambda^{VA}_i \mathcal{O}^{VA}_i\,,
 \end{equation}
and \cite{Kampf:2011ty,RuizFemenia:2003hm}
 \begin{equation}
\mathcal{L}_{2\,Res}^{odd}=\sum_{i=1}^3d_i\mathcal{O}_i^{VV}+\kappa_3^{PV}\mathcal{O}_3^{PV}\,.
 \end{equation}
The operators appearing in the two previous equations are ($h^{\mu\nu}=\nabla^\mu u^\nu+\nabla^\nu u^\mu$)
\begin{eqnarray}
\mathcal{O}_1^{PV}&=&i\langle[\nabla^\mu P,V_{\mu\nu}]u^\nu\rangle,\nonumber\\
\mathcal{O}_2^{PV}&=&i\langle[P,V_{\mu\nu}]f_-^{\mu\nu}\rangle;\nonumber\\
\mathcal{O}_1^{PA}&=&i\langle[P,A_{\mu\nu}]f_+^{\mu\nu}\rangle;\nonumber\\
\mathcal{O}_1^{VA}&=&\langle[V^{\mu\nu},A_{\mu\nu}]\chi_-\rangle,\nonumber\\
\mathcal{O}_2^{VA}&=&i\langle[V^{\mu\nu},A_{\nu\alpha}]h_\mu^\alpha\rangle,\nonumber\\
\mathcal{O}_3^{VA}&=&i\langle[\nabla^\mu V_{\mu\nu},A^{\nu\alpha}]u_\alpha\rangle,\nonumber\\
\mathcal{O}_4^{VA}&=&i\langle\nabla^\alpha V_{\mu\nu},A_\alpha^{\;\nu}]u^\mu\rangle,\\
\mathcal{O}_5^{VA}&=&i\langle[\nabla^\alpha V_{\mu\nu},A^{\mu\nu}]u_\alpha\rangle;\nonumber\\
\mathcal{O}_1^{VV}&=&\varepsilon_{\mu\nu\rho\sigma}\langle\left\{V^{\mu\nu},V^{\rho\alpha}\right\}\nabla_\alpha u^\sigma\rangle,\nonumber\\
\mathcal{O}_2^{VV}&=&i\varepsilon_{\mu\nu\rho\sigma}\langle\left\{V^{\mu\nu},V^{\rho\sigma}\right\}\chi_-\rangle,\nonumber\\
\mathcal{O}_3^{VV}&=&\varepsilon_{\mu\nu\rho\sigma}\langle\left\{\nabla_\alpha V^{\mu\nu},V^{\rho\alpha}\right\}u^\sigma\rangle,\nonumber\\
\mathcal{O}_3^{PV}&=&\varepsilon_{\mu\nu\alpha\beta}\langle\lbrace V^{\mu\nu},f_+^{\alpha\beta}\rbrace P\rangle.\nonumber
\end{eqnarray}
There is only one relevant operator with three resonance fields in either parity sector 
 \begin{equation}
     \mathcal{L}_{3\,Res}=i\lambda^{VAP}\langle[V_{\mu\nu},A^{\mu\nu}]P\rangle+\kappa^{PVV}\varepsilon_{\mu\nu\rho\sigma}\langle V^{\mu\nu}V^{\alpha\beta}P \rangle.
 \end{equation}
 
 Operators with a higer number of resonant fields will not be included, since otherwise one has to include subleading diagrams with loops where some of the internal lines are given by resonances. The present analysis is restricted to tree-level diagrams, which should already capture the leading effects associated with resonance exchange. One-loop diagrams with resonances are expected to be a numerically small correction since these would be subleading in the $1/N_C$ expansion \cite{Guevara:2015pza}. Such corrections will be neglected due to the already sizeable number of parameters involved in the tree-level analysis and the current precision of the experimental data.
 
 \subsection{Form Factors}\label{sec:Form_Factors}

 In this section we quote our results for the different contributions to the $F_V$, $F_A$, $A_2$ ,$A_4$ form factors, for $P=\pi,K$. All resonance propagators are to be understood as provided with an energy-dependent width ($M_R^2-x\to M_R^2-x-iM_R\Gamma_R(x)$, $x=W^2,k^2$) computed within $R\chi T$, using those in refs. \cite{GomezDumm:2000fz} ($\rho(770)$), \cite{Jamin:2006tk} ($K^*(892)$) and \cite{Dumm:2009va, Nugent:2013hxa} ($a_1(1260)$, including the $KK\pi$ cuts \cite{Dumm:2009kj}). A constant width will suffice for the very narrow $\omega(782)$ and $\phi(1020)$ mesons (their PDG \cite{Zyla:2020zbs} values will be taken). For the $K_1(1270/1400)$ states we will follow $\cite{Guo:2008sh}$.
 
 \begin{figure}[!ht]
     \centering
     \includegraphics[scale=0.3]{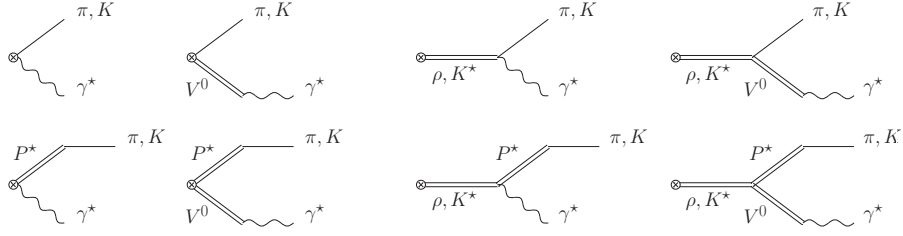}
     \caption{Feynman diagrams contributing to the vector part of the left hadronic current. {The circled cross vertex indicates vector current.} The resonance $P^\star$ is the pseudoscalar resonance corresponding to $\pi
     (1300)\equiv \pi'$ ($K(1460)\equiv K'$) for 
     $P=\pi(K)$. The resonance $V^0$ means $\omega$ for $P=\pi$ and $\rho^0,\omega,\phi$ for $P=K$.}
     \label{fig:MV_diagrams}
 \end{figure}

The vector form factors are ($N_C=3$ in QCD)
 
 \begin{equation}
     F_V^{(\pi)}(W^2,k^2)=\frac{1}{3f}\left\{-\frac{N_C\frac{}{}}{8\pi^2}+64 m_\pi^2 C_7^{W\star}-8C_{22}^{W}(W^2+k^2)\hspace*{10ex}
          \right.\nonumber\end{equation}\begin{equation}
         \hspace*{10ex}+\frac{4{F_V^{ud}}^2}{M_\rho^2-W^2}\frac{d_3(W^2+k^2)+d_{123}^\star m_\pi^2}{M_\omega^2-k^2}    \nonumber\end{equation}\begin{equation}
\hspace*{14ex}+\frac{2\sqrt{2}F_V^{ud}}{M_V}\frac{c_{1256}W^2-c_{1235}^\star m_\pi^2-c_{125}k^2}{M_\rho^2-W^2}
          \nonumber\end{equation}\begin{equation}
          \left.\hspace*{14ex}+\frac{2\sqrt{2}F_V^{ud}}{M_V}\frac{c_{1256}k^2-c_{1235}^\star m_\pi^2-c_{125}W^2}{M_\omega^2-k^2}
          \right\},
 \end{equation}
 \begin{equation}
   F_V^{(K)}(W^2,k^2)=\frac{1}{f}\left\{-\frac{N_C}{24\pi^2}+\frac{64}{3}m_K^2C_7^{W\star}+32C_{11}^W\Delta_{K\pi}^2-\frac{8}{3}C_{22}^W(W^2+k^2)
   \right.\nonumber\end{equation}\begin{equation}
   +\frac{2F_V^{us}\left[d_3(W^2+k^2)+d_{123}^\star m_K^2\right]}{M_{K^\star}^2-W^2}\left(\frac{F_V^{ud}}{M_\rho^2-k^2}+\frac{1}{3}\frac{F_V^{ud}}{M_\omega^2-k^2}-\frac{2}{3}\frac{F_V^{ss}}{M_\phi^2-k^2}\right)
   \nonumber\end{equation}\begin{equation}
   +\frac{2\sqrt{2}F_V^{us}}{3M_V}\frac{c_{1256}W^2-c_{1235}^\star m_K^2-c_{125}k^2+24c_4\Delta_{K\pi}^2}{M_{K^\star}^2-W^2}
   \end{equation}\begin{equation}
   \left.+\frac{\sqrt{2}\left(c_{1256}k^2-c_{1235}^\star m_K^2-c_{125}W^2\right)}{M_V}\left(\frac{F_V^{ud}}{M_\rho^2-k^2}+\frac{1}{3}\frac{F_V^{ud}}{M_\omega^2-k^2}-\frac{2}{3}\frac{F_V^{ss}}{M_\phi^2-k^2}\right)\right\},\nonumber
 \end{equation}
where $\Delta_{K\pi}^2=m_K^2-m_\pi^2$ and we have used the combinations of coupling constants \cite{Dumm:2012vb} 
 \begin{eqnarray}
 &&c_{125}=c_1-c_2+c_5,\nonumber\\
 &&c_{1256}=c_1-c_2-c_5+2c_6,\nonumber\\
 &&c_{1235}=c_1+c_2+8c_3-c_5,\\
 &&d_{123}=d_1+8d_2-d_3.\nonumber
 \end{eqnarray}
 $F_V^{uD,ss}$ and starred coefficients absorb $U(3)$ breaking contributions induced by $\lambda_V$ in eq. (\ref{eq:1R_V_even}) and pseudoscalar resonances, respectively.  Their expressions are given at the end of this subsection, after eq.~(\ref{eq22}).\\
 
 \begin{figure}[!ht]
     \centering
     \includegraphics[scale=0.28]{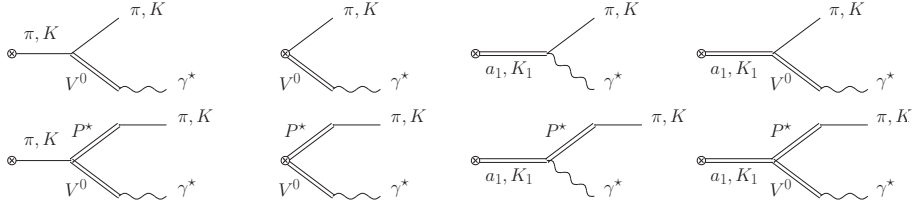}
     \caption{Feynman diagrams contributing to the axial part of the left hadronic current. {The circled cross vertex indicates axial current.}
     Conventions for $P^\star$ is the same as in the previous figure, the resonance $V^0$ means $\rho^0$ for $P=\pi$ and $\rho^0,\omega,\phi$ for $P=K$.}
     \label{fig:MA_diagrams}
 \end{figure}
 
 The axial form factors are~\footnote{We note two mistakes in writing $F_A^{(\pi)}$ in ref. \cite{Guevara:2013wwa}, see Appendix \ref{sec:Appendix_MA}. The result written here agrees with the one in ref. \cite{Guo:2010dv} for $k^2\to0$.}
 
 \begin{equation}
     F_A^{(\pi)}(W^2,k^2)= \frac{F_V^{ud}}{2f}\frac{F_V^{ud}-2G_V-m_\pi^2\frac{4\sqrt{2}d_m}{M_{\pi'}^2}(\lambda_1^{PV}+2\lambda_2^{PV})}{M_\rho^2-k^2}
   \nonumber\end{equation}\begin{equation}\label{eq:FApi}
  -\frac{F_A}{2f}\frac{F_A-2m_\pi^2\frac{4\sqrt{2}d_m}{M_{\pi'}^2}\lambda_1^{PA}}{M_{a_1}^2-W^2}+\frac{\sqrt{2}}{f}\frac{F_AF_V^{ud}}{M_{a_1}^2-W^2}\frac{\lambda_0^\star m_\pi^2-\lambda'k^2-\lambda''W^2}{M_\rho^2-k^2},
 \end{equation}
 
 \begin{equation}\label{eq:FAK}
     F_A^{(K)}(W^2,k^2)=-\frac{F_A}{2f}\frac{F_A-2m_K^2\frac{4\sqrt{2}d_m}{M_{K'}^2}\lambda_1^{PA}}{M_{K_1}^2-W^2}
   \nonumber\end{equation}\begin{equation}
   +\left[\frac{\sqrt{2}F_A}{2f}\frac{\lambda^\star_0m_K^2-\lambda'k^2-\lambda''W^2}{M_{K_1}^2-W^2}+\frac{F_V^{us}\left(F_V^{us}-2G_V+m_K^2\frac{4\sqrt{2}d_m}{M_{K'}^2}(\lambda_1^{PV}+2\lambda_2^{PV})\right)}{4f}\right]
   \nonumber\end{equation}\begin{equation}
   \times\left(\frac{F_V^{ud}}{M_\rho^2-k^2}+\frac{1}{3}\frac{F_V^{ud}}{M_\omega^2-k^2}+\frac{2}{3}\frac{F_V^{ss}}{M_\phi^2-k^2}\right),
 \end{equation}
 
 \begin{equation}
     A_2^{(\pi)}(W^2,k^2)=\hspace*{66ex}\nonumber\end{equation}\begin{equation}\label{eq:A2pi}\frac{2}{f}\left(G_V+\frac{2\sqrt{2}m_\pi^2d_m}{M_{\pi'}^2}\lambda_1^{PV}+\frac{\sqrt{2}F_A}{M_{a_1}^2-W^2}W^2(\lambda'+\lambda'')\right)\frac{F_V^{ud}}{M_\rho^2-k^2},
 \end{equation}
 
 \begin{equation}
     A_2^{(K)}(W^2,k^2)=\left(\frac{G_V}{f}+\frac{2\sqrt{2}m_K^2d_m}{M_{K'}^2}\frac{\lambda_1^{PV}}{f}+\frac{\sqrt{2}F_A}{M_{K_1}^2-W^2}\frac{W^2(\lambda'+\lambda'')}{f}\right)\nonumber\end{equation}\begin{equation}\hspace*{28ex}\times\left(\frac{F_V^{ud}}{M_\rho^2-k^2}+\frac{1}{3}\frac{F_V^{ud}}{M_\omega^2-k^2}+\frac{2}{3}\frac{F_V^{ss}}{M_\phi^2-k^2}\right),
 \end{equation}
 
 \begin{equation}\label{eq:A4pi}
     A_4^{(\pi)}(W^2,k^2)=\frac{2}{f}\frac{F_V^{ud}}{M_\rho^2-k^2}\left[\frac{G_V}{W^2-m_\pi^2}+\frac{2\sqrt{2}d_mm_\pi^2\lambda_1^{PV}}{M_{\pi'}^2\left(W^2-m_\pi^2\right)}+\frac{\sqrt{2}F_A(\lambda'+\lambda'')}{M_{a_1}^2-W^2}\right],
 \end{equation}

 \begin{equation}
     A_4^{(K)}(W^2,k^2)=\frac{1}{f}\left(\frac{G_V}{W^2-m_K^2}+\frac{2\sqrt{2}d_mm_K^2\lambda_1^{PV}}{M_{K'}^2\left(W^2-m_\pi^2\right)}+\frac{\sqrt{2}F_A(\lambda'+\lambda'')}{M_{K_1}^2-W^2}\right)\nonumber\end{equation}\begin{equation}\hspace*{28ex}\times\left(\frac{F_V^{ud}}{M_\rho^2-k^2}+\frac{1}{3}\frac{F_V^{ud}}{M_\omega^2-k^2}+\frac{2}{3}\frac{F_V^{ss}}{M_\phi^2-k^2}\right),\label{eq22}
 \end{equation}
 
 It is worth to notice that by replacing the $P$ propagator in $A_2^{(P)}$ and $A_4^{(P)}$ with the massless pole propagator, one recovers the linear dependence between both form factors, thus getting a congruent expression with those in reference \cite{Cirigliano:2006hb}. Therefore, the short-distance constraints obtained in this reference can be used as shown there if the Weinberg's sum rules are imposed. We, however, do not make use of these sum rules, as $F_{V/A}$ are fitted to data (see discussion in sections \ref{sec:SD_Constraints} and \ref{sec:Fit}).\\
 
We introduced the short-hand notation
\begin{eqnarray}
F_V^{ud}\equiv F_V+8m_\pi^2\lambda_V,\nonumber\\
F_V^{us}\equiv F_V+8m_K^2\lambda_V,\\
F_V^{ss}\equiv F_V+8(2m_K^2-m_\pi^2)\lambda_V,\nonumber
\end{eqnarray}
for the shifts appearing also in \cite{Guevara:2018rhj}. \footnote{As mentioned in section \ref{sec:Relavant_operators}, a similar shift can be introduced in $F_A$, however, the operator responsible for such shift can be absorbed through axial resonance field redefinitions \cite{Cirigliano:2006hb}.}\\

We also used \cite{GomezDumm:2003ku}
\begin{eqnarray}&&\hspace*{-2ex}-\sqrt{2}\lambda_0^\star=4\lambda_1^\star+\lambda_2+\frac{\lambda_4}{2}+\lambda_5,\nonumber\\ &&\sqrt{2}\lambda'=\lambda_2-\lambda_3+\frac{\lambda_4}{2}+\lambda_5\end{eqnarray} and \begin{equation}\hspace*{-3ex}\sqrt{2}\lambda''=\lambda_2-\frac{\lambda_4}{2}-\lambda_5.\nonumber\end{equation}
                                                                                                                                                                                                                                            

We employed several starred coefficients including $U(3)$ breaking contributions, as given below:

\begin{eqnarray}&&\lambda_1^\star=\lambda_1-\frac{\lambda^{VAP}d_m}{M_P^2},\nonumber\\
&&C_7^{W\star}=C_7^W+\frac{\kappa_5^Pd_m}{M_P^2},\nonumber\\
&&c_3^\star=c_3+\frac{\kappa_3^{PV}d_mM_V}{M_P^2}\label{eq:def_c3star},\end{eqnarray} implying
\begin{eqnarray}&&c_{1235}^\star=c_1+c_2+8c_3^\star-c_5,\;\mathrm{and}\label{eq:def_c1235star}\\
&&d_2^\star=d_2+\frac{\kappa^{VVP}d_m}{2M_P^2},\end{eqnarray} yielding
\begin{equation}d_{123}^\star=d_1+8d_2^\star-d_3.\end{equation}

We have first shown here the correction to $\lambda_1$ appearing in $\lambda_1^\star$, while the remaining starred couplings were already introduced in ref. \cite{Guevara:2018rhj}. \\

We will follow the
scheme explained in ref. \cite{Guo:2008sh} to account for the mixing between the $K_1(1270)=K_{1L}$ and the $K_1(1400)=K_{1H}$ states. This amounts to replacing, in eq. (\ref{eq:FAK}), $(M_{K_1}^2-W^2)^{-1}\to\mathrm{cos}^2\theta_A(M_{K_{1H}}^2-W^2)^{-1}+\mathrm{sin}^2\theta_A(M_{K_{1L}}^2-W^2)^{-1}$, with mixing angle $\theta_A\in[37,58]^\circ$.

\subsection{Short-distance constraints}\label{sec:SD_Constraints}
We will demand that the different form factors have an asymptotic behaviour in agreement with QCD \cite{Lepage:1980fj, Brodsky:1973kr}. Specifically, we will require their vanishing for large $\lambda$ in the $\lim_{\lambda\to\infty}F_V(\lambda W^2,0)$ and $\lim_{\lambda\to\infty}F_V(\lambda W^2,\lambda k^2)$ cases. We will do this first in the chiral limit and then at $\mathcal{O}(m_P^2)$~\footnote{Since we are considering a complete basis of chiral symmetry breaking operators at order $m_P^2$, we neglect higher-order chiral corrections.}, paralleling the discussion in ref. \cite{Guevara:2018rhj} for the neutral pseudoscalar transition form factors. In this way, we find the following relations:
\begin{itemize}
\item $F_V^{(\pi)}(W^2,k^2)$, $\mathcal{O}(m_P^0)$:
\begin{eqnarray}
C_{22}^W&=&0\,,\label{eq:SD_V_chiral_1}\\
c_{125}&=&0\,,\\
c_{1256}&=&-\frac{N_C M_V}{32\sqrt{2}\pi^2F_V}\,,\\
d_3&=&-\frac{N_CM_V^2}{64\pi^2F_V^2}\,.\label{eq:SD_V_chiral_last}
\end{eqnarray}
\item $F_V^{(\pi)}(W^2,k^2)$, $\mathcal{O}(m_P^2)$:
\begin{eqnarray}
\lambda_V&=&-\frac{64\pi^2F_V}{N_C}C_7^{W\star}\,,\\
c_{1235}^\star&=&\frac{N_CM_Ve^V_m}{8\sqrt{2}\pi^2F_V}+\frac{N_CM_V^3\lambda_V}{4\sqrt{2}\pi^2F_V^2}.\label{eq:SD_c1235star}
\end{eqnarray}
\item $F_V^{(K)}(W^2,k^2)$, $\mathcal{O}(m_P^0)$:
Same constraints as for the $\pi$ case, since both form factors~\footnote{$F_A^{P}$ and $A_{2,4}^{P}$ form factors are also identical in this limit for $P=\pi$ or $K$,  obviously.} are identical in the $U(3)$ symmetry limit.
\item $F_V^{(K)}(W^2,k^2)$, $\mathcal{O}(m_P^2)$:
\begin{eqnarray}
C_{11}^W&=&\frac{N_C\lambda_V}{64\pi^2F_V}\,.
\end{eqnarray}
\end{itemize}
For the sake of predictability and in order to further constrain the parameters in the form factor, we use the $VVP$ Green function, $\Pi_{VVP}(r^2,p^2,q^2)$, constraints \cite{Kampf:2011ty} obtained from the high-energy behaviour when $r^2\to\infty,p^2\to\infty$ and $q^2\to\infty$ and matching to the Operator Product Expansion (OPE) leading terms in the chiral and large-$N_C$ limits. These give 
\begin{eqnarray}
&&c_{125}=c_{1235}=0,\hspace*{25ex}c_{1256}=-\frac{N_CM_V}{32\sqrt{2}\pi^2F_V},\nonumber\\
&&\kappa_5^P=0,\hspace*{17ex}d_3=-\frac{N_CM_V^2}{64\pi^2F_V^2}+\frac{F^2}{8F_V^2}+\frac{4\sqrt{2}d_m\kappa_3^{PV}}{F_V},\nonumber\\
&&C_7^W=C_{22}^W=0,\hspace*{33ex}d_{123}=\frac{F^2}{8F_V^2}. \label{eq:SD_OPE_V} 
 \end{eqnarray}

Notice that these constraints coincide with our expressions in eqs. (\ref{eq:SD_V_chiral_1}-\ref{eq:SD_V_chiral_last}) and that they imply 
\begin{eqnarray}
    C_{11}^W&=&\lambda_V=C_7^{W\star}=0,\nonumber\\
    d_{123}&=&\frac{F^2}{8F_V^2},\nonumber\\
    d_m\kappa_3^{PV}&=&\frac{N_CM_{\pi'}^2e_m^V}{64\sqrt{2}\pi^2F_V}.
\end{eqnarray}

{One can see that from the definition of $c_{1235}^\star$ (eqs. (\ref{eq:def_c3star}) and (\ref{eq:def_c1235star})) combined with the last expression and the short-distance constraints eqs. (\ref{eq:SD_V_chiral_last}), (\ref{eq:SD_c1235star}) and (\ref{eq:SD_OPE_V}) would imply a relation of $e_m^V$ in terms of $F$ and $M_{\pi'}$, namely
\begin{equation}\label{eq:fake_emV}
 e_m^V=-\frac{2\pi^2F^2}{N_CM_{\pi'}^2}
\end{equation}
however, we do not rely on this relation since comparison with previous phenomenology \cite{Cirigliano:2003yq,Guo:2009hi} shows that the absolute value of eq. (\ref{eq:fake_emV}) obtained for $f\approx92$ MeV and $M_{\pi'}=1.3$ GeV is an order of magnitude smaller than required by phenomenology.}\\

On the other hand, no relation among parameters of the axial form factors can be obtained by taking the infinite virtualities limit,  
since it already has the right asymptotic behavior. Instead, we will rely on the relations obtained using the $VAP$ Green Function\footnote{See, however, the discussion in section 6.2 of \cite{Cirigliano:2007ga} comparing these short-distance constraints to the results in refs. \cite{Moussallam:1997xx, Knecht:2001xc}.} $\Pi_{VAP}(p^2,q^2,(p+q)^2)$ \cite{Cirigliano:2004ue}  in an analogous manner to that done for the $\Pi_{VVP}(r^2,p^2,q^2)$ Green Function.  \\

We recall that the simultaneous analysis of the scalar form factor \cite{Jamin:2000wn,  Jamin:2001zq} and the SS-PP sum rules \cite{Golterman:1999au} yields $d_m=f/(2\sqrt{2})$. Additionally, notice that $A_{2,4}^{(P)}$ depend on $\lambda_1^{PV}$. In turn, $F_A^{(P)}$ depends on $\lambda_1^{PA}$ and $\lambda_1^{PV}+2\lambda_2^{PV}$.
The appropriate short-distance behaviour of the $VAP$ Green Function \cite{Cirigliano:2004ue} fixes all of them but $\lambda_0^\star$ or, in other words $\lambda^{VAP}$, as noted in ref. \cite{Cirigliano:2006hb}
\begin{eqnarray}
    \lambda_0=\frac{f^2}{4\sqrt{2}F_VF_A}, & & \lambda'=\frac{f^2+F_A^2}{2\sqrt{2}F_VF_A}, \nonumber\\
    \lambda''=-\frac{f^2+F_A^2-2F_VG_V}{2\sqrt{2}F_VF_A}, &&  d_m\lambda_1^{PV}=-\frac{f^2}{4\sqrt{2}F_V},\nonumber\\ d_m\lambda_2^{PV}=\frac{3f^2+2F_A^2-2F_V^2}{16\sqrt{2}F_V},& & d_m\lambda_1^{PA}=\frac{f^2}{16\sqrt{2}F_A}\,.
\end{eqnarray}

Despite the relation for $d_m$ from the scalar form factor and the SS-PP Green's function, notice that there is no need for one since $d_m$ always appears multiplied by one of the other parameters to be constrained. We will also make use of the constraint \cite{Ecker:1988te}
\begin{equation}
    F_VG_V=f^2.
\end{equation}

  In order to gain predictability, we will use the values of $d_{123}^\star$, $M_V$ and $e_m^V$ given for the best fit of reference \cite{Guevara:2018rhj}, namely (their correlations are given in the quoted reference)
  \begin{eqnarray}
      d_{123}^{\star}&=&-(2.3\pm1.5)\times10^{-1},\nonumber\\
      M_V&=&(791\pm6)\mathrm{\hspace*{1ex}MeV},\\
      e_m^{V}&=&-(0.36\pm0.10).\nonumber
  \end{eqnarray}

\section{Phenomenological analysis}\label{sec:Pheno_anal}

\subsection{Phase space}\label{sec:phase_sapce}

{In order to compare our results with those of ref. \cite{Guevara:2013wwa} we use the same phase space configuration.
}We recall that the variables in ref. \cite{Guevara:2013wwa} are the invariant mass squared of the pseudo Goldstone and the neutrino, $s_{12}=m_{P\nu_\tau}^2$, the invariant mass squared of the charged lepton pair, $s_{34}=m_{\ell\overline{\ell}}^2$, two polar angles ${\theta_1},\,{\theta_3}$  and one azimuthal angle $\phi_3$, with the integration limits given by
\begin{subequations}\label{eq:PS_old}
\begin{align}
 (m_3+m_4)^2\leq &s_{34}\leq (M - m_1 - m_2)^2,\\
(m_1+m_2)^2\leq &s_{12}\leq (M - \sqrt{s_{34}})^2,\\
0\leq\theta_{1,3}\leq\pi,&\qquad0\leq\phi_3\leq2\pi.
\end{align}
\end{subequations}
If we identify the particle with tag 1 with $\nu_\tau$, the invariant mass of the weak gauge boson can be related to the Lorentz invariants of eqs. (\ref{eq:PS_old}) via
\begin{equation}\label{eq:W2}
   W^2=M_{234}^2=M^2+m_1^2-\frac{(M^2+s_{12}-s_{34})(s_{12}+m_1^2-m_2^2)}{2s_{12}}-X\beta_{12}\cos{\theta_1}\,,
\end{equation}
where $\beta_{ij}=\lambda^{1/2}(s_{ij},m_i^2,m_j^2)/s_{ij}$ and $X=\lambda^{1/2}(M^2,s_{12},s_{34})/2$, being $\lambda(a,b,c)=a^2+b^2+c^2-2ab-2ac-2bc$, the K\"all\'en lambda function. {Eq. (\ref{eq:W2}) allows us to eliminate $\theta_1$ in favor of $M_{234}$. The importance of the phase space configuration with $W^2$ instead of $\theta_1$ relies on the need to compute the $m_{\pi e^+e^-}$ spectrum in order to fit unconstrained parameters to the Belle invariant mass spectrum \cite{Jin:2019goo}.}
The kinematic limits on the non-angular variables for this phase space configuration read
\begin{subequations}\label{eq:Kinematical_limits}
 \begin{align}
    m_2+m_3+m_4&\leq\hspace*{1ex} M_{234}\hspace*{1ex}\leq M,\\
    (m_3+m_4)^2&\leq\hspace*{1ex} s_{34}\hspace*{1ex}\leq \left(M_{234}-m_2\right)^2,\\
    s_{12}^-&\leq\hspace*{1ex} s_{12} \hspace*{1ex}\leq s_{12}^+,
 \end{align}
\end{subequations}
where
\begin{eqnarray}
    s_{12}^\pm&=&    M^2+m_1^2+m_2^2+s_{34}-W^2+\frac{(M^2-m_1^2)(m_2^2-s_{34})}{W^2}\nonumber\\&&\pm\frac{1}{2W^2}\lambda^{1/2}(M^2,W^2,m_1^2)\lambda^{1/2}(W^2,s_{34},m_2^2)\,.
\end{eqnarray}
With this, the differential decay rate is given as 
\begin{eqnarray}\label{eq:dec_width}
    d\Gamma(\tau^-\to\nu_\tau P^-\ell\overline{\ell})
    &=&\frac{X\beta_{12}\beta_{34}}{4(4\pi)^6m_\tau^3}\overline{\left|\mathcal{M}\right|^2}ds_{34}ds_{12}d(\cos{\theta_1})d(\cos{\theta_3})d\phi_3\nonumber\\
    &=&\frac{\beta_{34}}{4
(4\pi)^6m_\tau^3}
\overline{\left|\mathcal{M}\right|^2}dM_{P\ell\overline{\ell}}^2
    ds_{34}ds_{12}d(\cos{\theta_3})d\phi_3\,.
\end{eqnarray}
\subsection{Fit to data}\label{sec:Fit}

   Short-distance QCD behaviour \cite{Guevara:2018rhj,Cirigliano:2006hb} does not constrain all parameters. Thus, we fit some of the remaining unknowns using the invariant mass spectra of the $W$ boson, $\invmass$, measured by the Belle collaboration \cite{Jin:2019goo}. We start with a four parameters fit ($F_V$, $F_A$, $\lambda_0$ and $\mB$, the branching fraction, are floated). 
   Despite the fact that the whole $\invmass$ spectrum has been measured, not all the data is available for this minimization since points below $\invmass<1.05$ GeV were used as a control region to validate the Monte Carlo simulation, leaving thus the most sensitive part to SD contributions as the signal region \cite{Jin:2019goo}; therefore we use for the minimization the data above the cut $\invmass=1.05$ GeV. 

We use only the data set for the $\tau^-\to\pi^-e^+e^-\nu_\tau$ mode~\footnote{This is the one shown in the plots of ref. \cite{Jin:2019goo}. We have checked better agreement with the Monte Carlo simulation (based on our previous paper \cite{Guevara:2013wwa}, see also \cite{Shekhovtsova:2012ra, Antropov:2019ald}) for this mode with respect to its charge-conjugated mode.}.  Comparison of this with the expected signal events distribution in this reference allows us to roughly quantify the deconvolution of signal from detector, which we ignore. We have assumed this to be an energy-independent effect for simplicity, and taken it into account as a systematic uncertainty in the data. This error turns out to be comparable to the one reported by Belle for the branching fraction measured above the cut. In addition to this, the Belle collaboration used the expressions of our ref. \cite{Guevara:2013wwa}; which however had typos in some of the  $F_A(t,k^2)$ terms {(see Appendix \ref{sec:Appendix_MA})}. Besides, trying to keep our previous analysis as simple as possible, it resulted  incomplete in the sense of $VAP$ Green's function analysis\footnote{Pseudoscalar resonance exchange and $\mO(p^6)$ operators in the $A_2^{(\pi)}$ and $A_4^{(\pi)}$ form factors are lacking in our ref. \cite{Guevara:2013wwa}. This leads to relating both form factors to the $\pi$ electromagnetic form factor \cite{Bijnens:1992en}.}, which altogether could lead to biased estimations of the decay observables. Both reasons motivate our choice of fitting the total branching fraction, $\mB$, as an additional parameter instead of simply computing it from the decay width expression in eq. (\ref{eq:dec_width}).\\
   
   We used the relation 
   \begin{equation}
       \int d\invmass \frac{1}{\Gamma}\frac{d\Gamma}{d\invmass}=1=
       \int d\invmass \frac{1}{N}\frac{dN}{d\invmass}=\sum_{\mathrm{bins}}\frac{1}{N}\frac{N_\mathrm{bin}}{\Delta\invmass},
   \end{equation}
where $N$ is the total number of events, $N_\mathrm{bin}$ is the number of events in a given bin and $\Delta\invmass$ is the bin width. We thus minimize the $\chi^2$ given by
\begin{equation}
    \chi^2 = \left(\frac{N\hspace*{1ex}\Delta\invmass}{\Gamma\hspace*{1ex}\varepsilon_\mathrm{bin}}\frac{d\Gamma}{d\invmass}-\frac{N_\mathrm{bin}}{\varepsilon_\mathrm{bin}}\right)^2+\left(\frac{\mathcal{B}-BR}{\varepsilon_{BR}}\right)^2    ,
\end{equation}
where $\varepsilon_\mathrm{bin}$ is the experimental uncertainty in a given bin, $BR$ is the branching ratio for $\invmass\geq1.05$ GeV reported by Belle \cite{Jin:2019goo}, $\varepsilon_{BR}$ its error and $\mathcal{B}$ the one obtained integrating our differential decay width above this cut. We recall that $BR(\tau^- \to \nu_{\tau}\pi^-e^+e^-)=(5.90 \pm0.53\pm0.85\pm0.11) \times10^{-6}$ \cite{Jin:2019goo}, where the first uncertainty is statistical, the second is systematic, and the third is due to the model dependence. In our fits, we have first realized that, varying all four parameters, there happen to be many quasidegenerate best fits and that the correlations among the fit  parameters ($F_V$, $F_A$, $\lambda_0$ and $\mB$) are always large. We interpret this as the data not being precise enough to disentangle the physical solutions among all close to the global minimum. Then, we proceeded to further simplify the fits by making constant one of these four parameters (not $\mathcal{B}$, as our systematic error due to unfolding is comparable to the overall uncertainty of $BR$).\\

We present two sets of fits as our reference results. One fixing $F_A=130$ MeV ($\sim\sqrt{2}F$, in agreement with \cite{Roig:2013baa}), and the other setting  $\lambda_0^\star=102\times10^{-3}$ (which is in the ballpark of previous estimates for $\lambda_0$, and neglects the contribution of pseudoscalar resonances to the starred coupling, see \cite{Miranda:2020wdg} and refs. therein). Considering individually the $\pi^+$ or $\pi^-$ sets we find that, apart from the incompatibility among both sets in several bins, the $\pi^+$ data set leads to the unphysical condition $F_A>F_V$. Also, fixing $F_V$ to its short-distance prediction $F_V\sim\sqrt{3}F\sim159$ MeV \cite{Roig:2013baa}, yields fits with larger $\chi^2$ that we disregard.
One way of interpreting this feature would be that excited resonances (at least the $\rho(1450)$ state and its interference with the $\rho(1700)$ resonance) are needed for an improved description of the data. However, given the errors of the measurement and the lack of $m_{e^+e^-}$ invariant mass distribution data we are not able to test such more sophisticated theory input, which introduces several additional parameters that remain unconstrained after applying the short-distance conditions (see, e. g. ref. \cite{Mateu:2007tr}). We thus understand that our fitted values of $F_V$ are effectively capturing missing dynamics in our description (such as the $\rho(770)$ excitations).\\

\begin{figure}[!ht]
\centering\includegraphics[scale=0.75]{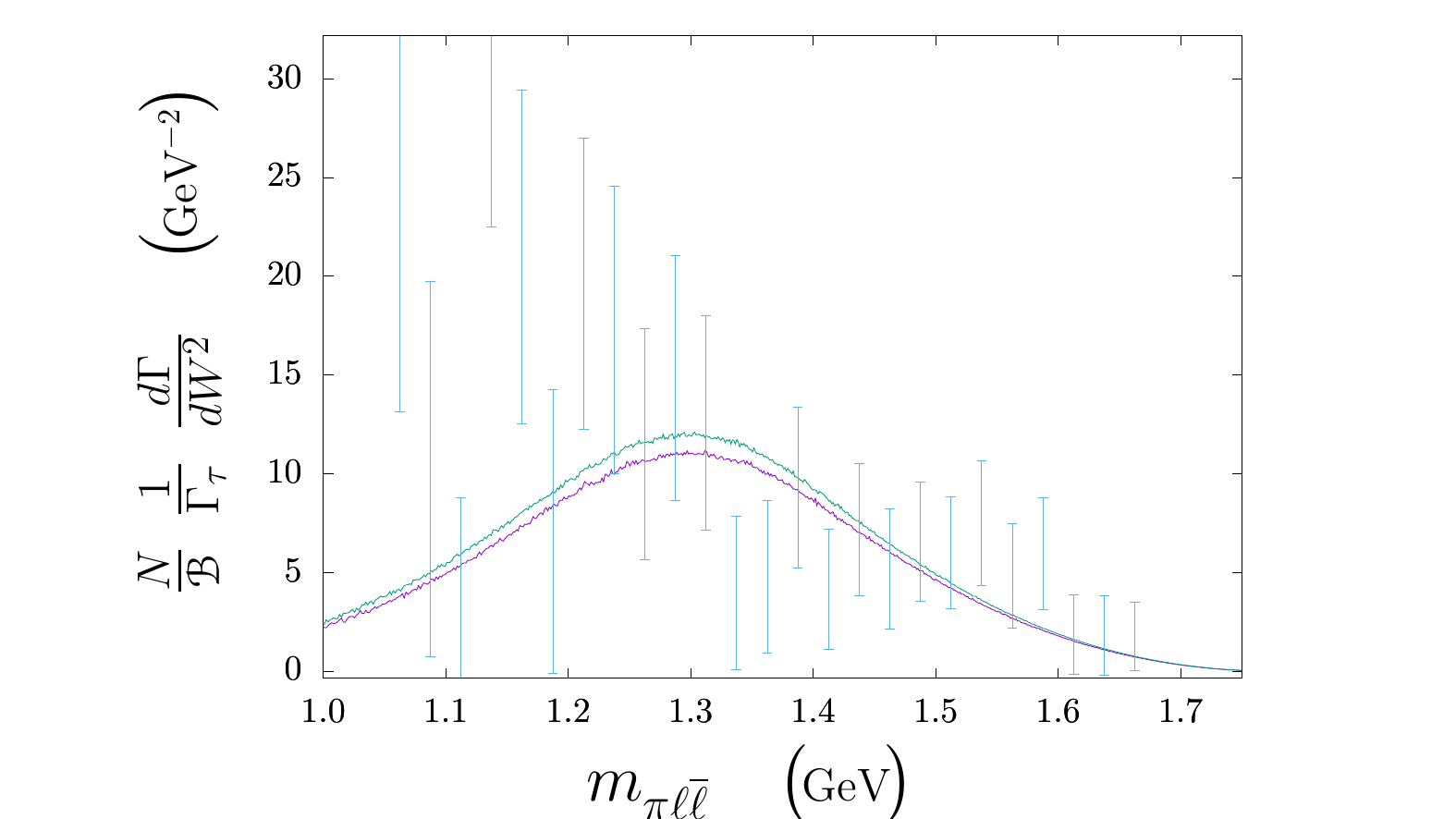}
\caption{Normalized invariant mass spectra obtained with the two sets of parameters obtained from fitting to the Belle data. The purple line corresponds to the data with $F_A$ fixed, while the green one stands for that with $\lambda_0^\star$ fixed. The blue data corresponds to measurements of $\tau^-$ decays, which show best agreement with our model. \cite{Jin:2019goo} }\label{fig:fit_spectra}
\end{figure}

{Considering the Weinberg sum rules \cite{Weinberg:1967kj}
\begin{subequations}
 \begin{align}
  \sum_i\left(F_{V_i}^2-F_{A_i}^2\right)&=f^2,\\
  \sum_i\left(F_{V_i}^2M_{V_i}^2-F_{A_i}^2M_{A_i}^2\right)&=0,
 \end{align}
\end{subequations}
and taking only the contributions from the lightest nonet of resonances (along with $M_A=\sqrt{2}M_V$), one finds that the fitted value for $F_V$ approaches the  prediction from these relations, namely that\footnote{Also found from short distance constrictions of vector and axial form factors considering only one multiplet \cite{Ecker:1988te,Ecker:1989yg}.} $F_V=\sqrt{2}f$. However, and as has been stated above, the consistent short distance limit when operators contributing to 2 and 3-point Green's Functions are considered should be \cite{Roig:2013baa} $F_V=\sqrt{3}f$. This makes us believe that the dynamics from heavier copies of the $\rho$ meson must be affecting the constraints on the decay constant $F_V$. Of course, these copies will undoubtedly affect the contribution to the chiral-order $p^4$ LECs of the non-resonant Lagrangian when integrating the resonances out, however, we still assume they approximately saturate them and neglect the operators of such Lagrangian.\footnote{Since the copies of the $\rho$ must have analogous dynamics, the operators contributing to the LECs of $\mathcal{O}(p^4)$ must be the same with the sustitution $\rho\to\rho'$ (and so on for heavier copies). Thus, their contributions must read
\begin{eqnarray}
 &&L_1^V=\sum_i\frac{G_{V_i}^2}{8M_{V_i}^2},\hspace*{10ex} L_9^V=\sum_i\frac{F_{V_i}G_{V_i}}{2M_{V_i}^2},\hspace*{10ex}L_{10}^V=-\sum_i\frac{F_{V_i}^2}{4M_{V_i}^2},
\end{eqnarray}
with $L_2^2=2L_1^V$ and $L_3^V=-6L_1^V$.
}
}

Our results are shown in table \ref{tab:fit_par}, the corresponding $\chi^2/dof\sim1.2$ is reasonably good and $\mathcal{B}$ is in agreement with the Belle data within less than 1 standard deviation in both cases. According to these results, we cannot exclude that pseudoscalar resonances give sizeable contributions to $\lambda_0^\star$. We consider both fit results in table \ref{tab:fit_par} as benchmarks for our predictions in the remainder of this work (we will refer to them as 'the two sets'). The difference among the corresponding two results can be taken as a first, rough estimate of our model-dependent error.

\begin{table}[!ht]
    \centering
    \begin{tabular}{c c c}\hline
    & set 1 & set 2\\
              &  \footnotesize $F_A=130$ MeV & \footnotesize $\lambda_0^\star=102\times10^{-3}$\\\hline\hline
        $F_A$ & 130 MeV& ($122\pm0$) MeV\\
        $F_V$ & (135.5$\pm1.1$) MeV& ($137.4\pm1.6$) MeV\\
        $\lambda_0^\star$ & $(384\pm 0)\times10^{-3}$ & 102$\times10^{-3}$\\
        $\mB$ &$(6.01\pm0)\times10^{-6}$ &$(6.36\pm0.12)\times10^{-6}$\\\hline
        $\chi^2/dof$ & 31.1/26& 31.4/26\\\hline\hline
    \end{tabular}
    \caption{Our best fit
    results for $F_A$ , $F_V$, $\lambda_0^*$ and $\mathcal{B}$. For the fit  results shown on the left (right) columns we fix $F_A=130$ MeV  ($\lambda_0^\star=102\times10^{-3}$),  respectively. A $0$ error means that the fit uncertainty in the parameter is negligible with respect to its central value.}
    \label{tab:fit_par}
\end{table}

\subsection{Predictions for the $\tau^-\to \nu_\tau P^- \ell \overline{\ell}$ decays}\label{sec:results}

\begin{table}[!ht]
\centering
\begin{tabular}{c c c |c}\hline
      &&\hspace*{-22ex}$\mB(\tau\to\nu_\tau P\ell\overline{\ell})$&\\
     $P,\ell$ & set 1 & set 2 & IB\\\hline\hline
     $\pi,e$&$(2.38\pm0.28\pm0.11)\cdot10^{-5}$&$(2.45\pm0.45\pm0.04)\cdot10^{-5}$& $1.457(5)\cdot10^{-5}$ \\
     $\pi,\mu$&$(8.45\pm2.45\pm1.09)\cdot10^{-6}$&$(9.15\pm3.25\pm0.25)\cdot10^{-6}$&$1.5935(4)\cdot10^{-7}$\\     $K,e$&$(1.17\pm0.26\pm0.09)\cdot10^{-6}$&$(1.11\pm0.28\pm0.04)\cdot10^{-6}$&$3.225(5)\cdot10^{-7}$\\
     $K,\mu$&$(6.4\pm1.9\pm0.8)\cdot10^{-7}$&$(5.85\pm1.75\pm0.20    )\cdot10^{-7}$&$3.4191(8)\cdot10^{-9}$\\\hline
\end{tabular}\caption{Branching ratios for the different $\tau$ decay channels. In the middle columns, our prediction for the full branching ratio accounting for both (dominant)  systematic and statistical uncertainties (see main text). In the right column we show the SI contribution with the error arising from numerical integration of the differential decay width. }\label{tab:BR}
\end{table}

 By generating 2400 points in the parameter space making a Gaussian variation of parameters, taking into account the correlations among them, we computed the sum of the SD and the SD-SI interference contributions to the branching fractions for the full phase space. We also computed for the $P=\pi$ and $\ell=e$ channel the SI contribution to $\mB$ with the cut $\invmass\geq1.05$ GeV, 
 \begin{equation}\label{eq:BR_cut}
     \left.\frac{}{}\mB(IB)\right|_{\invmass\geq1.05\mathrm{ GeV}}=(1.599\pm0.003)\times10^{-7}.
 \end{equation}
As expected, it is a factor $\sim37$ smaller than the Belle measurement, which confirms $\invmass\geq1.05$ GeV is a good cut to study structure-dependent effects.\\

\begin{figure}[!ht]
\hspace*{-3ex}\includegraphics[scale=0.45]{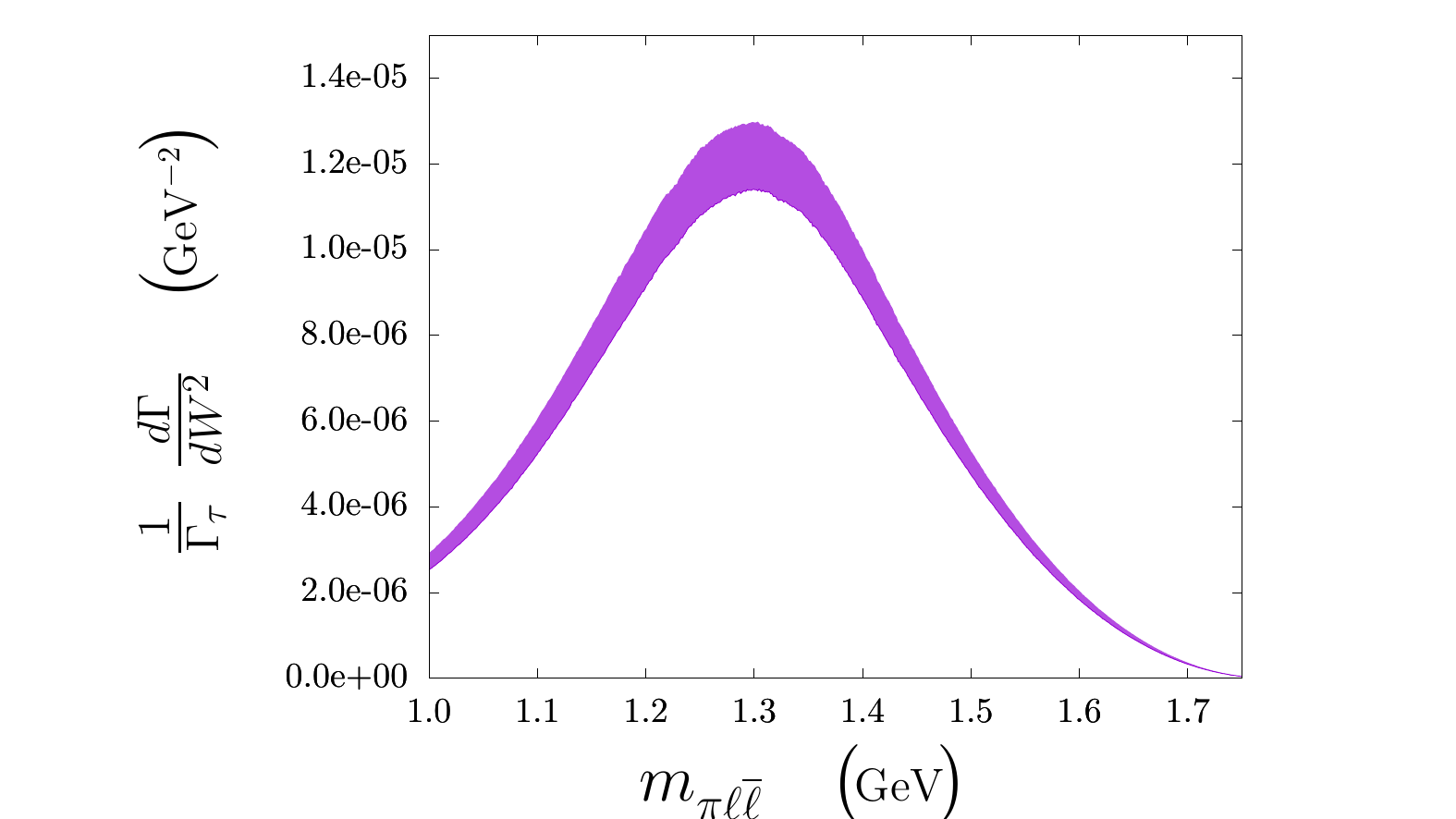}\hspace*{-6.5ex}\includegraphics[scale=0.45]{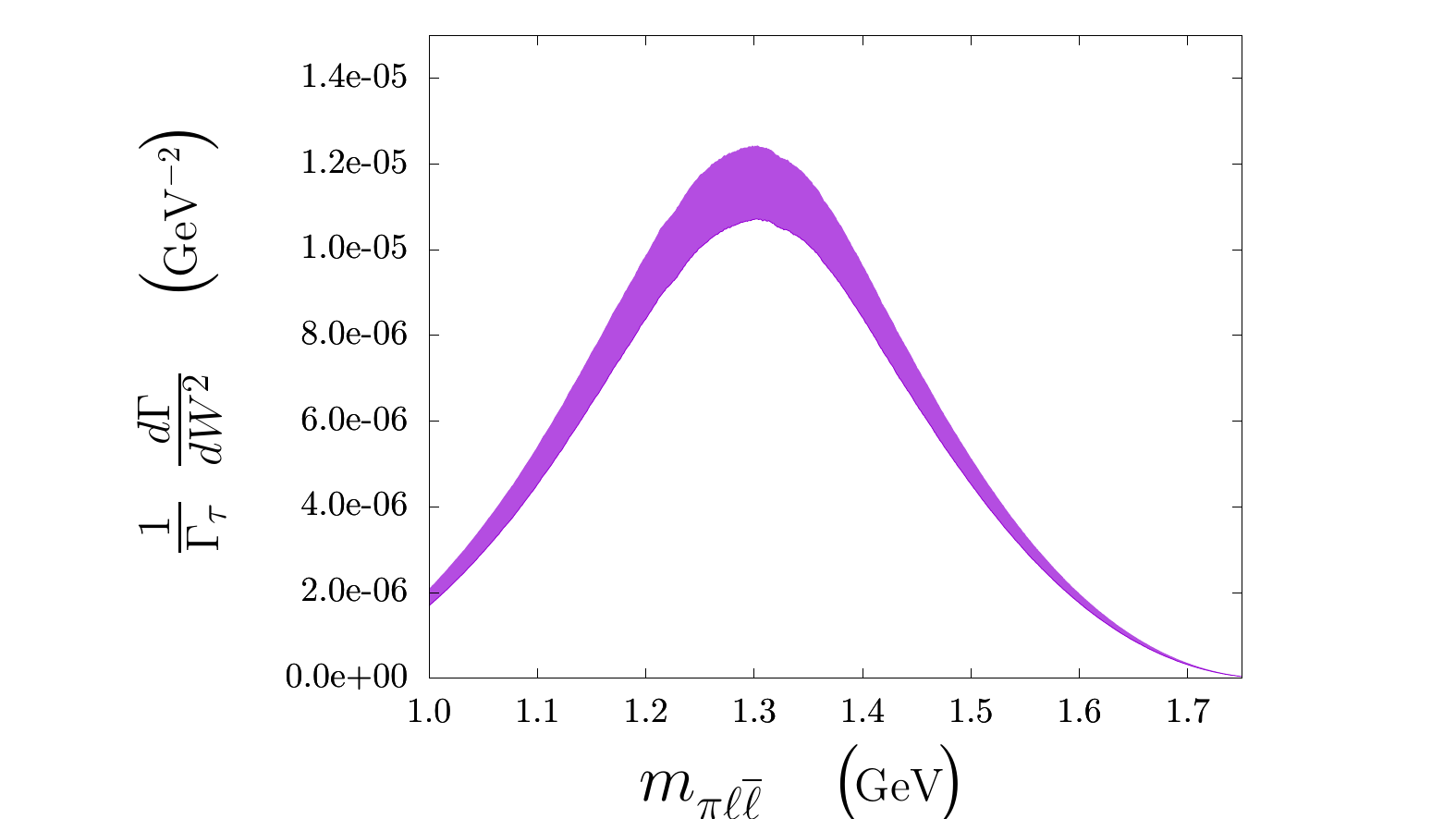}
\caption{Invariant mass spectra $\invmass$ for $P=\pi$, the thickness represents the error band obtained from the difference between the two sets. The plot in the left is for $\ell=e$, while the other is for $\ell=\mu$.
}\label{fig:W2_pi}
\end{figure}

  Thus, the $\mB$ adding the SI contribution gives the total branching ratios shown in table \ref{tab:BR}, where the first (dominant) error includes the uncertainty from unfolding and from the difference between $\mathcal{B}$ and BR and the second error was obtained from the Gaussian distribution of the fitted parameters. Also, in the same table, we show the SI contributions to the branching fractions for the different decay channels in the complete phase space.\\
 
\begin{figure}[!ht]
\hspace*{-3ex}\includegraphics[scale=0.45]{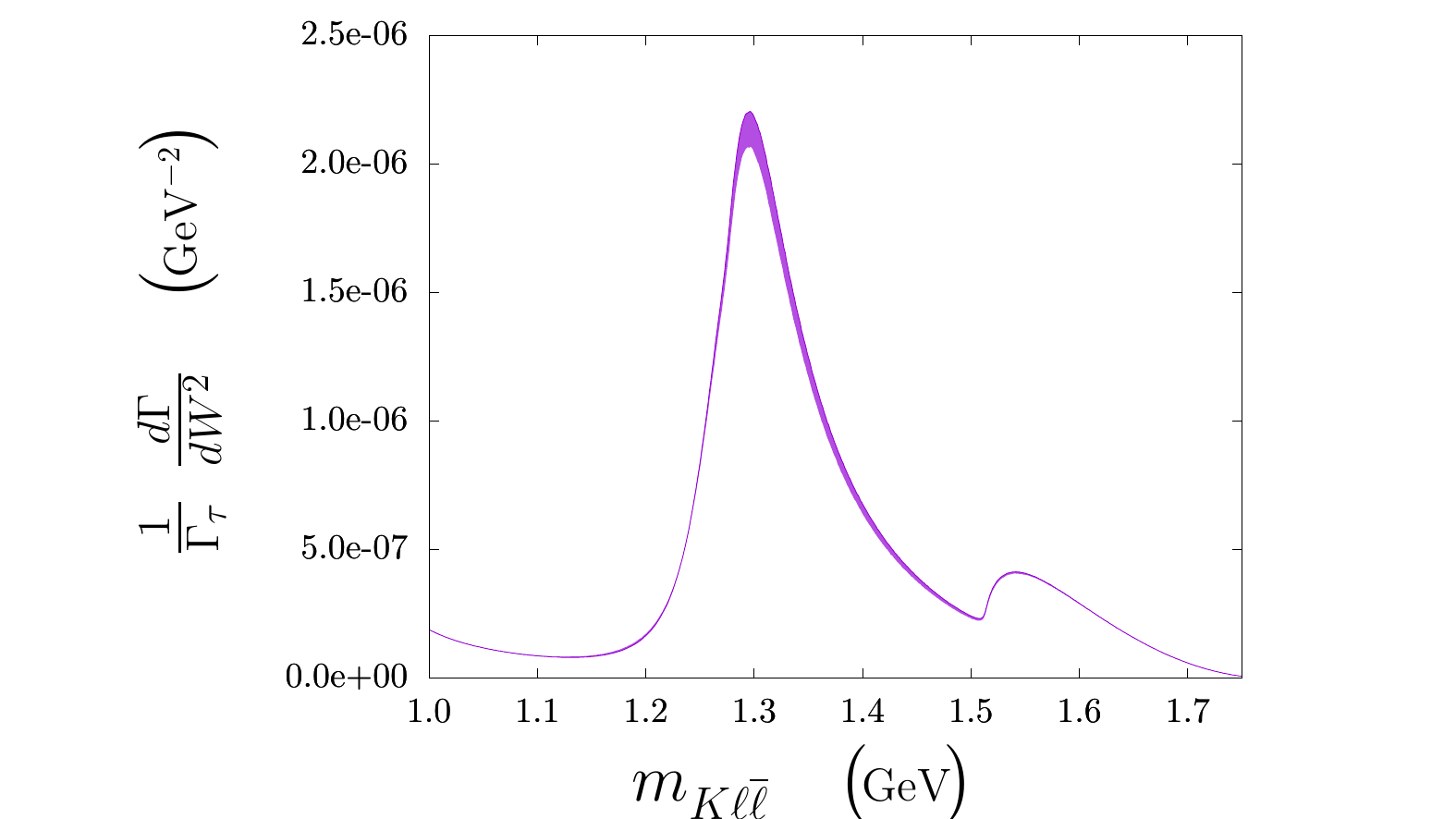}\hspace*{-6.5ex}\includegraphics[scale=0.45]{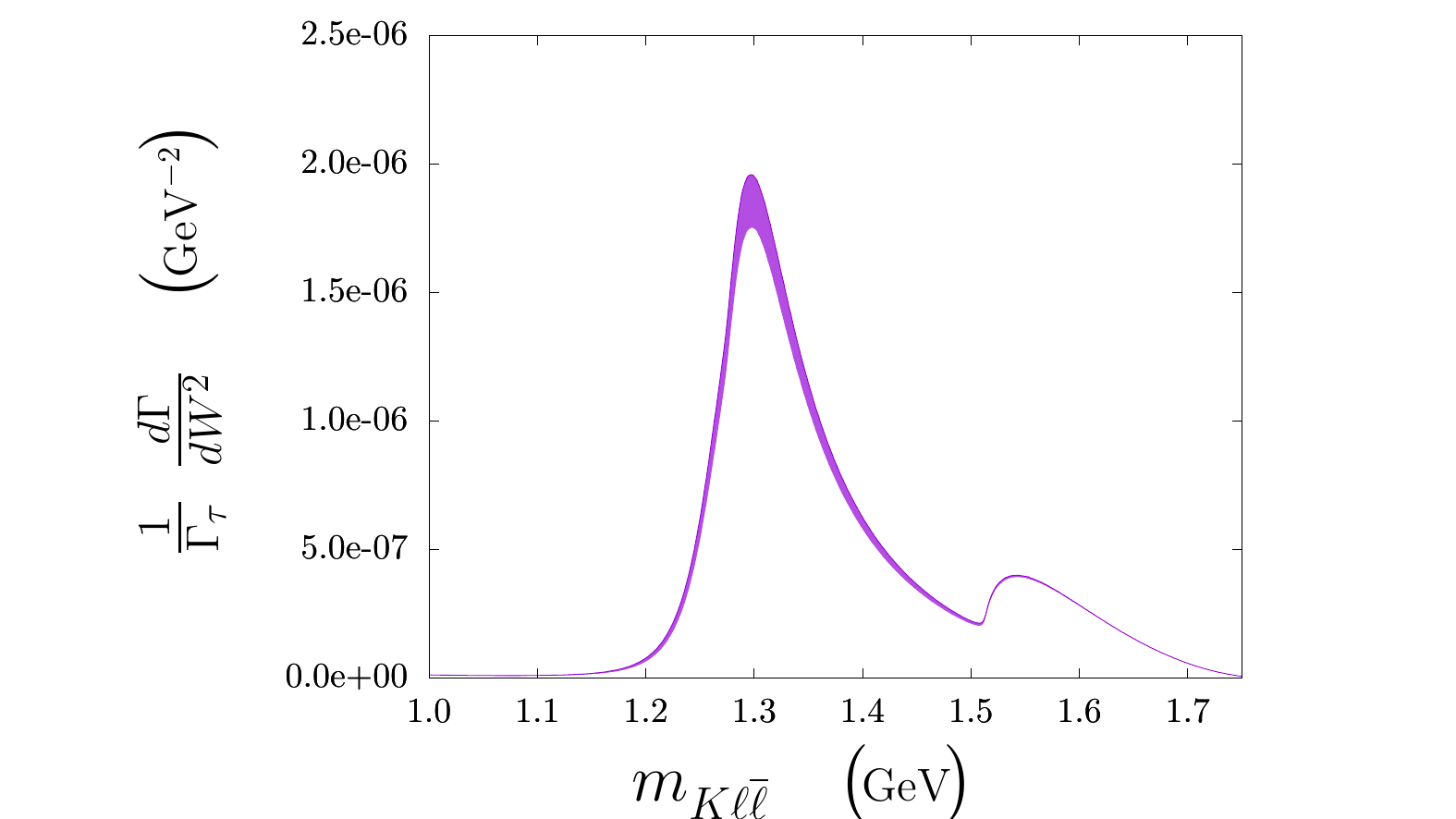}
\caption{Same as Figure \ref{fig:W2_pi} for $P=K$.
}\label{fig:W2_K}
\end{figure}
 
\begin{figure}[!ht]
\hspace*{-3ex}\includegraphics[scale=0.45]{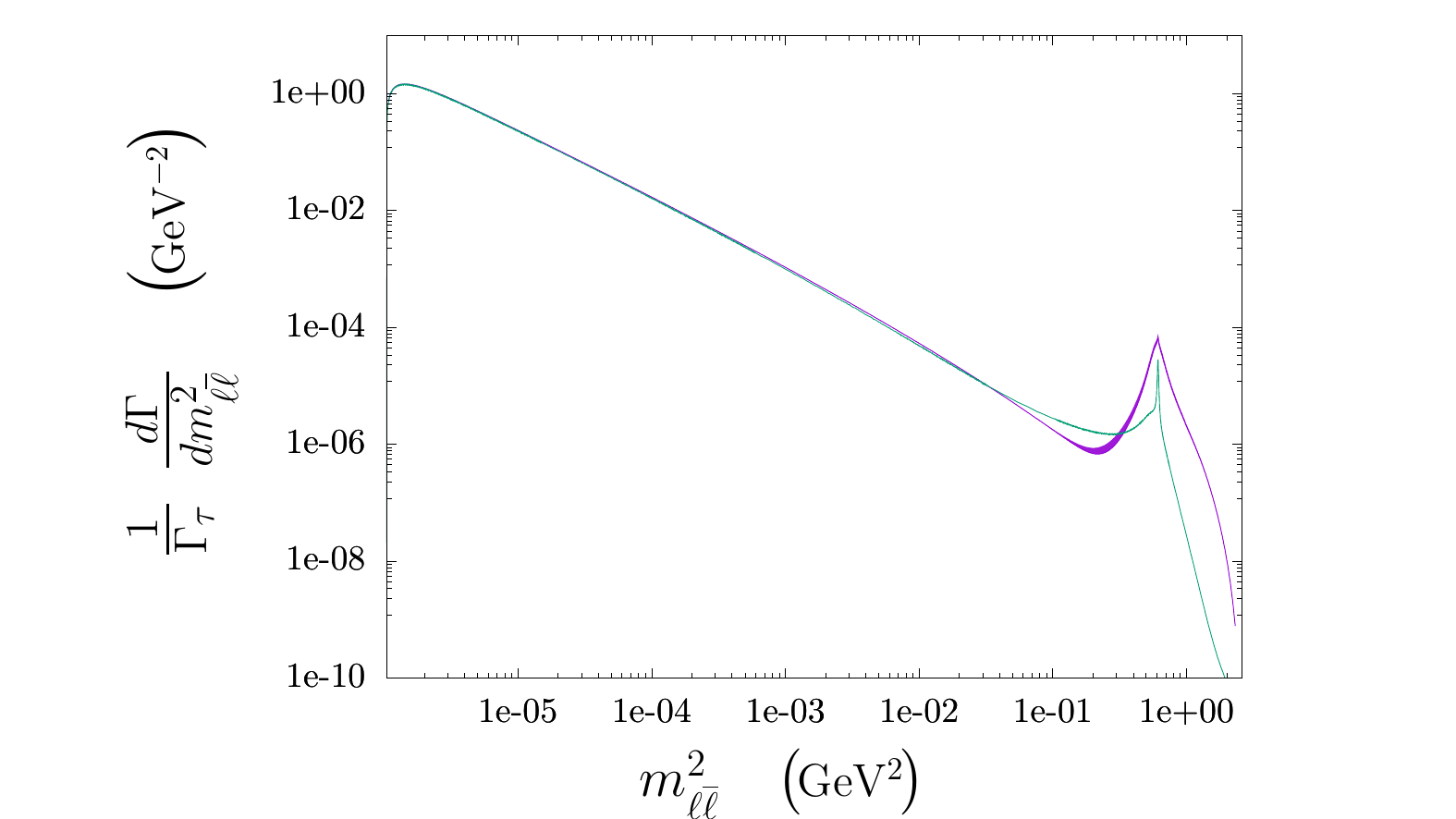}\hspace*{-6.5ex}\includegraphics[scale=0.45]{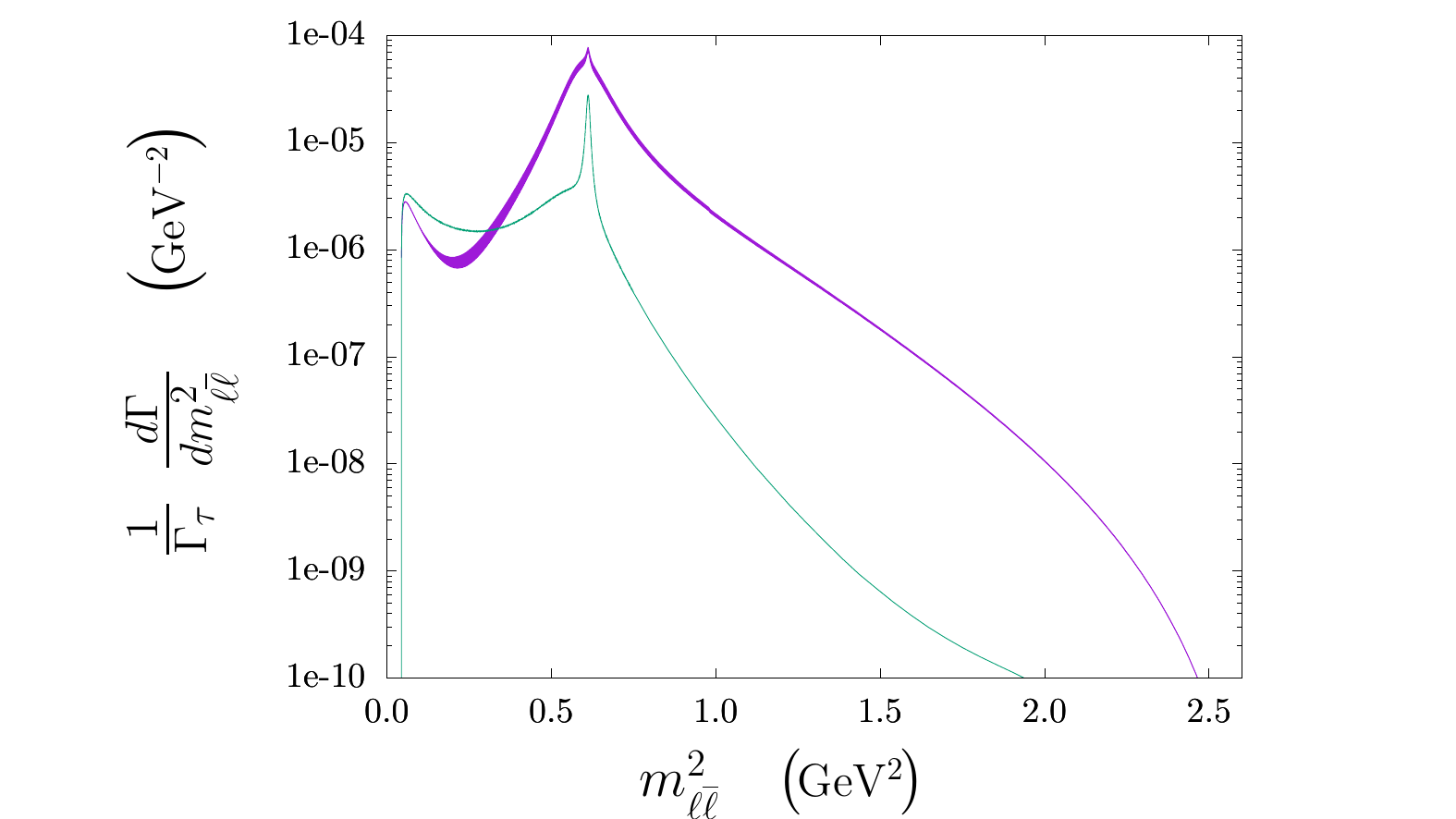}
\caption{Invariant mass spectra $m_{\ell\overline{\ell}}$ for $P=\pi$, the thickness of the purple line represents the error band obtained from the difference between the two sets. The green line is the prediction of ref. \cite{Guevara:2013wwa}. The left-hand plot is for $\ell=e$, while the other is for $\ell=\mu$
.}\label{fig:s34_pi}
\end{figure}

\begin{figure}[!ht]
\hspace*{-3ex}\includegraphics[scale=0.45]{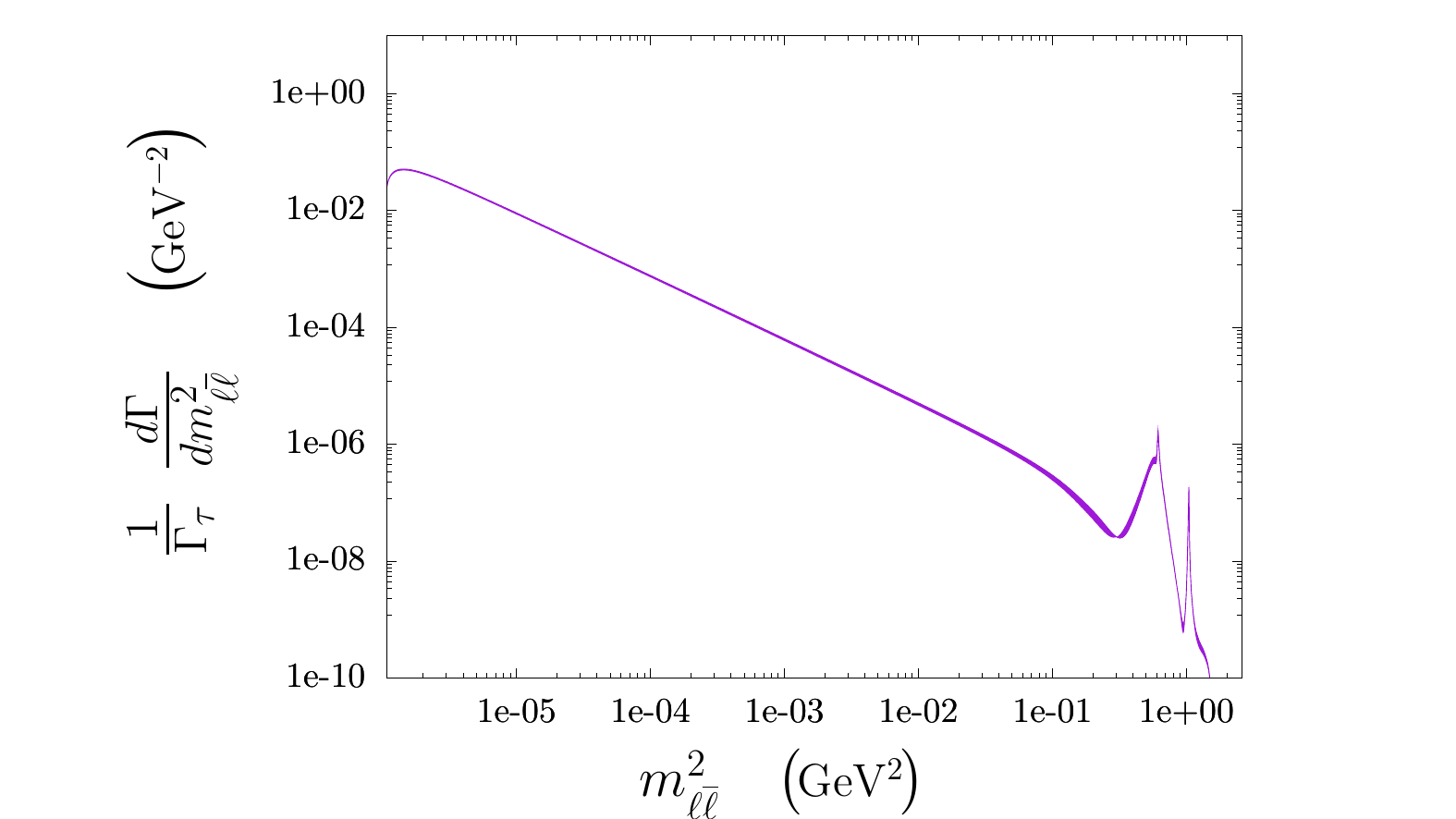}\hspace*{-6.5ex}\includegraphics[scale=0.45]{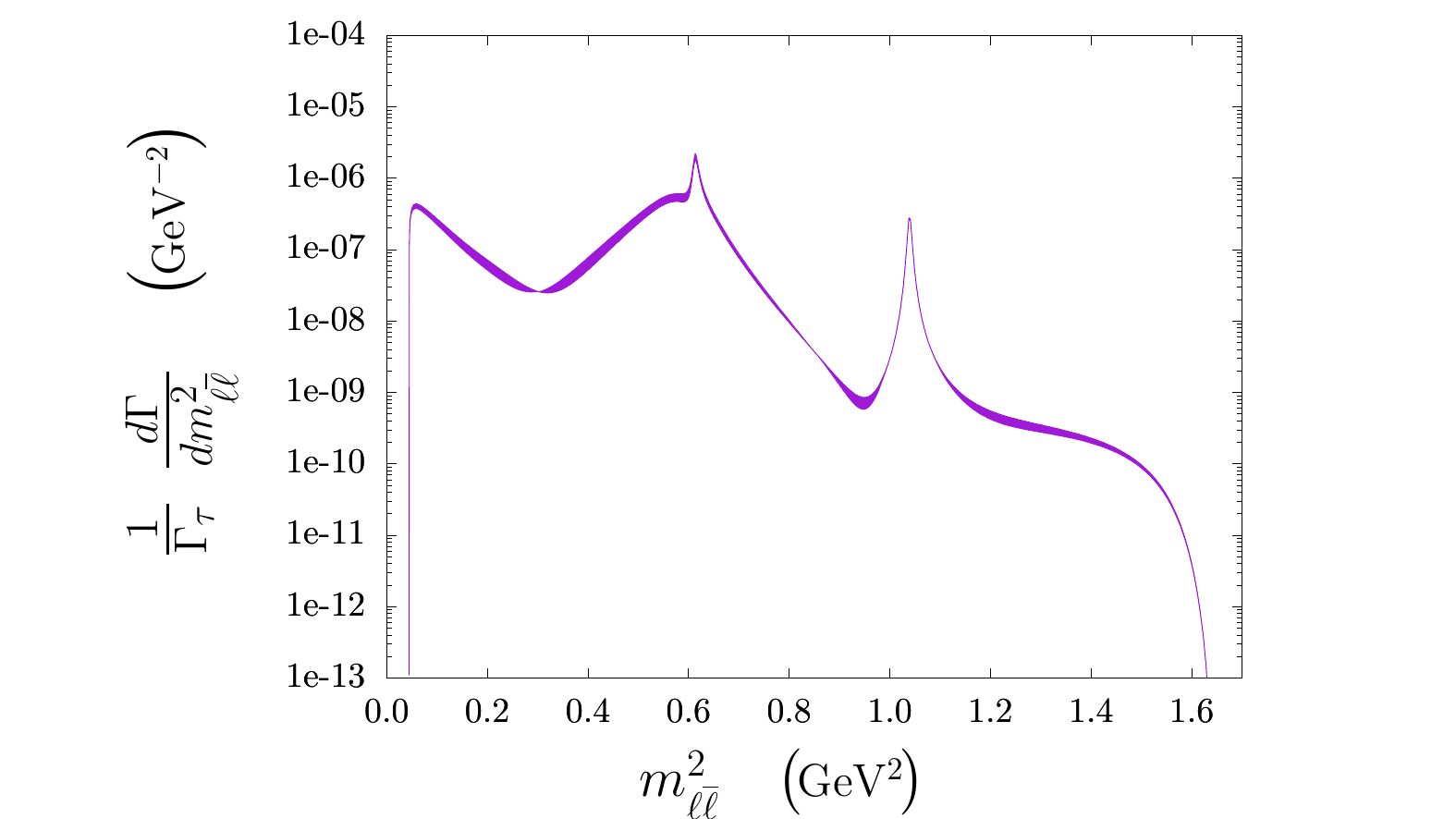}
\caption{Invariant mass spectra $m_{\ell\overline{\ell}}$ for $P=K$, the thickness of the purple line represents the error band obtained from the difference between the two sets. The plot on the left is for $\ell=e$, while the other is for $\ell=\mu$.
}\label{fig:s34_K}
\end{figure}

 From the discussion at the beginning of section \ref{sec:Fit} and from the results shown in table \ref{tab:BR}, we take the $\mB$ of each decay channel to be within the range obtained from the union of the intervals given by each set of fitted parameters, the latter ranges defined as the intervals given by each central value of Table \ref{tab:BR} and its uncertainties. Also, we computed the $W^2$ spectra for the different decay channels shown in Figures \ref{fig:W2_pi} and \ref{fig:W2_K} using both sets of fitted parameters, where the error band was obtained by taking the difference between them. The same was done for the dilepton spectra in Figures \ref{fig:s34_pi} and \ref{fig:s34_K}. Measurement of these observables at Belle-II \cite{Kou:2018nap} will be crucial for further reducing the uncertainties shown.

\section{Conclusions}\label{sec:concl}

 Motivated by recent measurements of the Belle Collaboration \cite{Jin:2019goo}, in  this paper we present an improved prediction of the  $\tau^-\to \pi^- e^+e^-\nu_\tau$ decay. 
This includes a more accurate description of the structure-dependent parts of the decay amplitudes, by taking into account  the first order flavor-breaking  corrections to the form factors involved in the $W\pi \gamma^*$ vertex. 
As done for the $VVP$ Green's function \cite{Roig:2013baa}, we found that the inclusion of the pseudoscalar resonance is needed in order to obtain compatible expressions between the $VAP$ Green's function and the form factors. 
In the high-energy limit we find that these expressions 
give the same constraints on the parameters of the resonance Lagrangians, as happens in the $VVP$ case.\\
 
 \begin{table}[!ht]
     \centering
     \begin{tabular}{cc}\hline
          $P,\ell$& $\mB(\tau^-\to\nu_\tau P^-\ell\overline{\ell})$ \\ \hline\hline
          $\pi,e$&$ (2.41\pm0.40\pm0.12)\cdot10^{-5}$ \\
          $\pi,\mu$&$ (9.15\pm3.25\pm1.12)\cdot10^{-6}$ \\
          $K,e$&$ (1.13\pm0.30\pm0.09)\cdot10^{-6}$ \\
          $K,\mu$&$ (6.2\pm2.1\pm0.8)\cdot10^{-7}$ \\\hline
     \end{tabular}
     \caption{Branching ratios for the different decay channels. The central value is the mean of the union of intervals given in both columns of Table \ref{tab:BR}, the first error covers the width of such union of ranges (see discussion below eq. (\ref{eq:BR_cut})) and the second error is the quadratic mean of statistical uncertainties in Table \ref{tab:BR}.}
     \label{tab:Final_BRs}
 \end{table}
 
 We have obtained a reasonably good fit of the parameters that remain unconstrained after applying the SD behaviour to the form factors. 
 A better set of data for  the invariant mass spectrum of the hadronic current would allow to determine physically meaningful parameters in a unique way
 . It is worth to recall that despite the fact that we had access to the $\tau^+\to \bar{\nu}_\tau\pi^+e^+e^-$ spectra obtained from Belle, we found some inconsistencies that make  
 unreliable the fits to data from both positive and negative tau decays 
 (see discussion in section \ref{sec:Fit}). We have therefore only considered the data set of the $\tau^-$ decays. From the results in Table \ref{tab:BR}, we conclude that our best result for the branching fractions is the union of ranges given for both fitting sets. This is shown in Table \ref{tab:Final_BRs}, where the central value is the mean of the union of these intervals. The results for the $P=\pi$ case agree with those in ref. \cite{Guevara:2013wwa}, where $(1.7^{+1.1}_{-0.3})\cdot10^{-5}$ for $\ell=e$ and $[3\cdot10^{-7},1\cdot10^{-5}]$ for $\ell=\mu$ were predicted. Thus, we have reached a 
 more precise determination of the branching ratios for the $\pi$ decay channels than the previous ones in ref. \cite{Guevara:2013wwa}. Also, 
 similar observables for the $\tau^\pm\to\nu_\tau K^\pm\ell\overline{\ell}$ channels are predicted for the first time
 .\\
 
 Despite the great 
 achievement of the Belle collaboration \cite{Jin:2019goo} discovering the $\tau^-\to\nu_\tau\pi^-e^+e^-$ decays
 , our study shows the need for better data (hopefully from Belle-II \cite{Kou:2018nap} and forthcoming facilities) in order to increase our knowledge of these decay modes. The  
 $\invmass$ spectrum shown in Figure \ref{fig:fit_spectra} is consistent with the destructive interference of the $\rho(1450)$ and $\rho(1700)$ resonances; 
 however, current data uncertainties prevent investigating the dynamics involved in the interplay of such resonances. The effect of $\rho$ excitations does not seem, however, negligible,  since by imposing the known behaviour \cite{Roig:2013baa} $F_V=\sqrt{3}F$ to the fit gives a far worse $\chi^2$ than those in Table \ref{tab:fit_par}, which are closer to $F_V=\sqrt{2}F$ (which holds with a minimal resonance Lagrangian beyond which we go in this and in our previous work on the subject). We assume that the effect of these heavier copies of the $\rho$ meson are responsible for this shift in the value of $F_V$. 

\section*{Acknowledgements}
We are indebted to Denis Epifanov and Yifan Jin for leading the Belle analysis of these decays, and measuring for the first time the $\tau\to\pi e^+e^- \nu_\tau$ decays. We specially acknowledge Yifan Jin for sharing with us detailed information on their study and providing us with the simulated Monte Carlo generation. We also aknowledge Pablo S\'anchez Puertas for usefull comments on short distance constraints. A.G. was supported partly by the Spanish MINECO and European FEDER funds (grant FIS2017-85053-C2-1-P) and Junta de Andaluc\'ia (grant FQM-225) and partly by the Generalitat Valenciana (grant Prometeo/2017/053). G.L.C. ackowledges funding from Ciencia de
Frontera Conacyt project No. 428218 and perfil PRODEP IDPTC 162336, and P.R. by the SEP-Cinvestav Fund (project number 142), grant PID2020-114473GB-I00 funded by MCIN/AEI/10.13039/501100011033 and by Cátedras Marcos Moshinsky (Fundación Marcos Moshinsky), that also supported A. G.

\appendix
\section{Previous and current expressions of axial form factors}\label{sec:Appendix_MA}

When we compare the expressions for the axial part of the decay amplitude we see that the following relations should be fulfilled
\begin{subequations}
\begin{align}
& F_A^\text{new}(W^2,k^2)=\frac{1}{2}F_A^\text{old}(t,k^2),\\
& A_2^\text{new}(W^2,k^2)\leftrightarrow\frac{1}{2}A_2^\text{old}(k^2),\\
& A_4^\text{new}(W^2,k^2)\leftrightarrow\frac{1}{2}A_4^\text{old}(k^2),
\end{align}
\end{subequations}
where $t=W^2$. There is, however, some mistakes in the form factors of reference \cite{Guevara:2013wwa}. There we have
\begin{eqnarray}\label{eq:new}
 &&F_A^\text{old}(t,k^2)=\frac{F_V^2}{F}\left(1-2\frac{G_V}{F_V}\right)D_\rho(k^2)-\frac{F_A^2}{F}D_{a_1}(t)\nonumber\\&&\hspace*{15ex}+\frac{F_AF_V}{\sqrt{2}F}D_\rho(k^2)D_{a_1}(t)(-\lambda''t+\lambda_0m_\pi^2),
\end{eqnarray}
where $D_\rho$ and $D_{a_1}$ are the propagators of the $\rho$ and $a_1$ respectively. However, when we neglect the contributions stemming from the $U(3)_V$-breaking contributions in eq. (\ref{eq:FApi}) we get
\begin{eqnarray}\label{eq:old_new}
 &&2F_A^\text{new}(t,k^2)=\frac{F_V^2}{F}\left(1-2\frac{G_V}{F_V}\right)D_\rho(k^2)-\frac{F_A^2}{F}D_{a_1}(t)\nonumber\\&&\hspace*{15ex}+{\color{red}2\sqrt{2}}\frac{F_AF_V}{F}D_\rho(k^2)D_{a_1}(t)(-\lambda''t{\color{red}-\lambda'k^2}+\lambda_0m_\pi^2),
\end{eqnarray}
where we show in red the factors and terms missing in the expression for $F_A^\text{old}(t,k^2)$ of ref. \cite{Guevara:2013wwa}.\\

Furthermore, since ref. \cite{Guevara:2013wwa} works in the chiral limit $A_2^\text{old}(t,k^2)$ and $A_4^\text{old}(t,k^2)$ are linearly dependent and are replace with the form factor $B$, such that
\begin{subequations}
 \begin{align}
  A_2^\text{old}(t,k^2)&\to-2B(k^2),\\
  A_4^\text{old}(t,k^2)&\to-\frac{2B(k^2)}{k^2+2p\cdot k}.
 \end{align}
\end{subequations}
As said previously, ref. \cite{Guevara:2013wwa} works on the chiral limit, therefore the $\lambda_1^{PV}$ term in eqs. (\ref{eq:A2pi}) and eqs. (\ref{eq:A4pi}) do not contribute. Nevertheless and for the sake of simplicity, ref. \cite{Guevara:2013wwa} gives $B$ in terms entirely of the $I=1$ part of the $\pi^+\pi^-$ vector form factor. This means that the contribution from the $a_1$ meson is being neglected.\\  

\begin{figure}[!ht]
 \centering\includegraphics[scale=0.75]{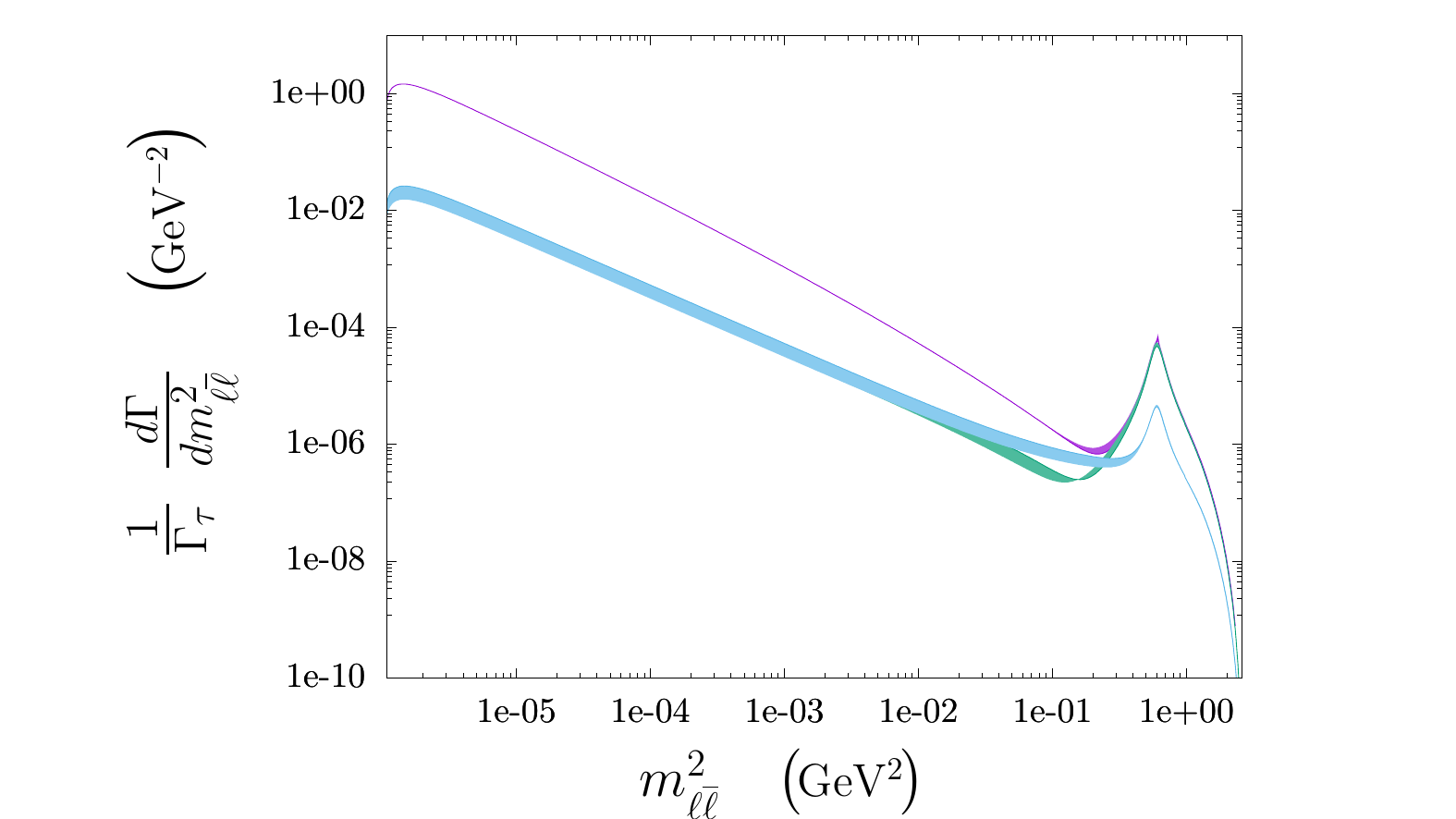}\caption{Invariant mass spectra in a double logarithmic scale for the complete amplitude (purple band), its contribution of the axial part (green band) and the same contribution but with the mistakes of ref. \cite{Guevara:2013wwa} (pale-blue band). The width of the band is an uncertainty computed as for Figures \ref{fig:s34_pi}}\label{fig:MA}
\end{figure}

\begin{figure}[!ht]
 \centering\includegraphics[scale=0.75]{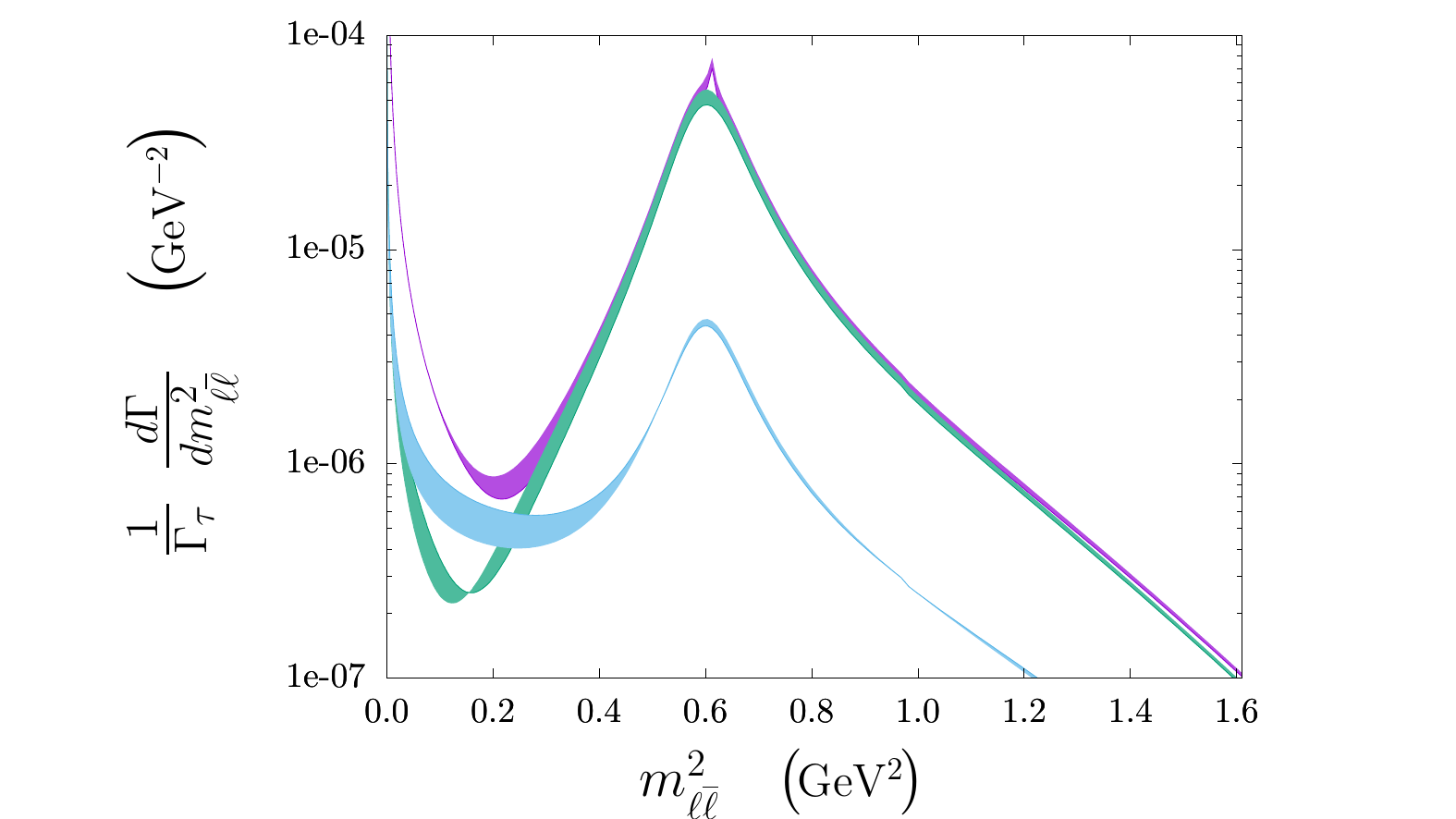}\caption{Invariant mass spectra of Figure \ref{fig:MA} using the same color code, with a logarithmic scale for the vertical axis, showing the region for their maximum contribution to $\mathcal{B}$. }\label{fig:MA_zoom}
\end{figure}

As a consequence of all this, we can see that the total spectrum gets really affected by such differences: We computed only the contribution of the axial amplitude to the $m_{e^+e^-}$ spectrum ({\it i.e.,} turning off the SI and vector contributions in eqs. (\ref{eq:amplitudes}) and keeping only that of eq. (\ref{eq:amplitudes_MA})), and computed also such spectrum but with the axial form factors of reference \cite{Guevara:2013wwa}. The complete spectrum is shown in Figure \ref{fig:MA} with a double logarithmic scale and a zoom with a logarithmic scale for the vertical axis in Figure \ref{fig:MA_zoom}.

As shown in ref. \cite{Guevara:2013wwa}, the importance of the SD contributions start at $m_{e^+e^-}^2\sim0.1\text{ GeV}^2$. Thus, as shown in Figures \ref{fig:MA} and \ref{fig:MA_zoom}, the total invariant mass spectrum (purple band) gets almost completely saturated by the axial contribution\footnote{at lower energies, the total invariant mass spectrum gets saturated by the SI contribution} (green band) at $m_{e^+e^-}^2\gtrsim0.1\text{ GeV}^2$. In these Figures we also show the spectrum obtained with the mistakes shown in eq. (\ref{eq:old_new}) (pale-blue band), keeping our constrained couplings and values for fitted parameters, we see that there is an important difference between the spectra with the new and old form factors. Their contribution to the branching fraction of the $\tau^-\to\nu_\tau\pi^-e^+e^-$ decay obtained from the spectra are an order of magnitude away
\begin{subequations}
 \begin{align}
  \left.\frac{}{}\mathcal{B}\right|_{IB,V\to0}^\text{new}&=1.09\times10^{-5},\\
  \left.\frac{}{}\mathcal{B}\right|_{IB,V\to0}^\text{old}&=1.66\times10^{-6},
 \end{align}
\end{subequations}
In contrast, we computed also the contribution from the axial amplitude to the branching ratio of the $\tau^-\to\nu_\tau\pi^-e^+e^-$ decay as done for the complete $\mathcal{B}$ in Table \ref{tab:BR} for the new form factors both, considering (fb) and neglecting (fc) the $U(3)_V$ breaking terms
\begin{subequations}
 \begin{align}
  \left.\frac{}{}\mathcal{B}\right|_{IB,V\to0}^\text{fb}&=(1.03\pm0.10)\times10^{-5},\\
  \left.\frac{}{}\mathcal{B}\right|_{IB,V\to0}^\text{fc}&=(1.02\pm0.10)\times10^{-5},
 \end{align}
\end{subequations}
showing thus that the difference between our results and those in ref. \cite{Guevara:2013wwa} stems from the differences between eqs. (\ref{eq:new}) and (\ref{eq:old_new}), and not from the $U(3)_V$ breaking terms.


\begin{thebibliography}{99}
\bibitem{Guevara:2016trs}
A.~Guevara, G.~López-Castro and P.~Roig,
Phys. Rev. D \textbf{95} (2017) no.5, 054015.

\bibitem{Husek:2017vmo}
T.~Husek, K.~Kampf, S.~Leupold and J.~Novotny,
Phys. Rev. D \textbf{97} (2018) no.9, 096013.

\bibitem{Kampf:2018wau}
K.~Kampf, J.~Novotn\'y and P.~Sánchez-Puertas,
Phys. Rev. D \textbf{97} (2018) no.5, 056010.

\bibitem{Guevara:2015pza}
A.~Guevara, G.~López Castro, P.~Roig and S.~L.~Tostado,
Phys. Rev. D \textbf{92} (2015) no.5, 054035.

\bibitem{Guevara:2013wwa}
P.~Roig, A.~Guevara and G.~López Castro,
Phys. Rev. D \textbf{88} (2013) no.3, 033007.

\bibitem{Bijnens:1992en}
J.~Bijnens, G.~Ecker and J.~Gasser,
Nucl. Phys. B \textbf{396} (1993), 81-118.

\bibitem{Jin:2019goo}
Y.~Jin \textit{et al.} [Belle],
Phys. Rev. D \textbf{100} (2019) no.7, 071101.

\bibitem{GutierrezSantiago:2020bhy}
J.~L.~Guti\'errez Santiago, G.~L\'opez Castro and P.~Roig,
Phys. Rev. D \textbf{103} (2021) no.1, 014027.


\bibitem{5leptons}
A. Flores-Tlalpa, G. Lopez Castro  and P. Roig, 
J. High Energ. Phys. 2016, 185 (2016). 

\bibitem{Arroyo-Urena:2021nil}
M.~A.~Arroyo-Ure\~na, G.~Hern\'andez-Tom\'e, G.~L\'opez-Castro, P.~Roig and I.~Rosell,
[arXiv:2107.04603 [hep-ph]], to be published in Phys. Rev. D, and work in progress.

\bibitem{Aoyama:2020ynm}
T.~Aoyama
\textit{ et al.}, The Muon g-2 Theory Initiative,
Phys. Rept. 887 (2020) 1-166.

\bibitem{Ecker:1988te}
G.~Ecker, J.~Gasser, A.~Pich and E.~de Rafael,
Nucl. Phys. B \textbf{321} (1989), 311-342.

\bibitem{Ecker:1989yg}
G.~Ecker, J.~Gasser, H.~Leutwyler, A.~Pich and E.~de Rafael,
Phys. Lett. B \textbf{223} (1989), 425-432.

\bibitem{Gasser:1983yg}
J.~Gasser and H.~Leutwyler,
Annals Phys. \textbf{158} (1984), 142.

\bibitem{Gasser:1984gg}
J.~Gasser and H.~Leutwyler,
Nucl. Phys. B \textbf{250} (1985), 465-516.

\bibitem{Weinberg:1978kz}
S.~Weinberg,
Physica A \textbf{96} (1979) no.1-2, 327-340.

\bibitem{Guevara:2018rhj}
A.~Guevara, P.~Roig and J.~J.~Sanz-Cillero,
JHEP \textbf{06} (2018), 160.

\bibitem{Roig:2014uja}
P.~Roig, A.~Guevara and G.~López Castro,
Phys. Rev. D \textbf{89} (2014) no.7, 073016.

\bibitem{Cirigliano:2006hb}
V.~Cirigliano, G.~Ecker, M.~Eidemüller, R.~Kaiser, A.~Pich and J.~Portolés,
Nucl. Phys. B \textbf{753} (2006), 139-177.

\bibitem{Knecht:2001xc}
M.~Knecht and A.~Nyffeler,
Eur. Phys. J. C \textbf{21} (2001), 659-678.

\bibitem{Cirigliano:2004ue}
V.~Cirigliano, G.~Ecker, M.~Eidemüller, A.~Pich and J.~Portolés,
Phys. Lett. B \textbf{596} (2004), 96-106.

\bibitem{Bernard:1991zc}
V.~Bernard, N.~Kaiser and U.~G.~Meissner,
Nucl. Phys. B \textbf{364} (1991), 283-320.

\bibitem{SanzCillero:2004sk}
J.~J.~Sanz-Cillero,
Phys. Rev. D \textbf{70} (2004), 094033.

\bibitem{Guo:2014yva}
Z.~H.~Guo and J.~J.~Sanz-Cillero,
Phys. Rev. D \textbf{89} (2014) no.9, 094024.

\bibitem{Wess:1971yu}
J.~Wess and B.~Zumino,
Phys. Lett. B \textbf{37} (1971), 95-97.

\bibitem{Witten:1983tw}
E.~Witten,
Nucl. Phys. B \textbf{223} (1983), 422-432.

\bibitem{Bijnens:2001bb}
J.~Bijnens, L.~Girlanda and P.~Talavera,
Eur. Phys. J. C \textbf{23} (2002), 539-544.

\bibitem{Kampf:2011ty}
K.~Kampf and J.~Novotny,
Phys. Rev. D \textbf{84} (2011), 014036.

\bibitem{Bijnens:1999sh}
J.~Bijnens, G.~Colangelo and G.~Ecker,
JHEP \textbf{02} (1999), 020.

\bibitem{Roig:2013baa}
P.~Roig and J.~J.~Sanz Cillero,
Phys. Lett. B \textbf{733} (2014), 158-163.

\bibitem{Kadavy:2020hox}
T.~Kadav\'y, K.~Kampf and J.~Novotny,
JHEP \textbf{10} (2020), 142.

\bibitem{RuizFemenia:2003hm}
P.~D.~Ruiz-Femenía, A.~Pich and J.~Portolés,
JHEP \textbf{07} (2003), 003.

\bibitem{GomezDumm:2003ku}
D.~Gómez Dumm, A.~Pich and J.~Portolés,
Phys. Rev. D \textbf{69} (2004), 073002.

\bibitem{Cirigliano:2003yq}
V.~Cirigliano, G.~Ecker, H.~Neufeld and A.~Pich,
JHEP \textbf{06} (2003), 012.

\bibitem{Guo:2009hi}
Z.~H.~Guo and J.~J.~Sanz-Cillero,
Phys. Rev. D \textbf{79} (2009), 096006.

\bibitem{GomezDumm:2000fz}
D.~Gómez Dumm, A.~Pich and J.~Portolés,
Phys. Rev. D \textbf{62} (2000), 054014.

\bibitem{Jamin:2006tk}
M.~Jamin, A.~Pich and J.~Portolés,
Phys. Lett. B \textbf{640} (2006), 176-181.

\bibitem{Dumm:2009va}
D.~G.~Dumm, P.~Roig, A.~Pich and J.~Portolés,
Phys. Lett. B \textbf{685} (2010), 158-164.

\bibitem{Nugent:2013hxa}
I.~M.~Nugent, T.~Przedzinski, P.~Roig, O.~Shekhovtsova and Z.~Was,
Phys. Rev. D \textbf{88} (2013), 093012.

\bibitem{Dumm:2009kj}
D.~G.~Dumm, P.~Roig, A.~Pich and J.~Portolés,
Phys. Rev. D \textbf{81} (2010), 034031.

\bibitem{Zyla:2020zbs}
P.~A.~Zyla \textit{et al.} [Particle Data Group],
PTEP \textbf{2020} (2020) no.8, 083C01.

\bibitem{Guo:2008sh}
Z.~H.~Guo,
Phys. Rev. D \textbf{78} (2008), 033004.

\bibitem{Dumm:2012vb}
D.~Gómez Dumm and P.~Roig,
Phys. Rev. D \textbf{86} (2012), 076009.

\bibitem{Guo:2010dv}
Z.~H.~Guo and P.~Roig,
Phys. Rev. D \textbf{82} (2010), 113016.

\bibitem{Lepage:1980fj}
G.~P.~Lepage and S.~J.~Brodsky,
Phys. Rev. D \textbf{22} (1980), 2157.

\bibitem{Brodsky:1973kr}
S.~J.~Brodsky and G.~R.~Farrar,
Phys. Rev. Lett. \textbf{31} (1973), 1153-1156.

\bibitem{Cirigliano:2007ga}
V.~Cirigliano and I.~Rosell,
JHEP \textbf{10} (2007), 005.

\bibitem{Moussallam:1997xx}
B.~Moussallam,
Nucl. Phys. B \textbf{504} (1997), 381-414.

\bibitem{Jamin:2000wn}
M.~Jamin, J.~A.~Oller and A.~Pich,
Nucl. Phys. B \textbf{587} (2000), 331-362.

\bibitem{Jamin:2001zq}
M.~Jamin, J.~A.~Oller and A.~Pich,
Nucl. Phys. B \textbf{622} (2002), 279-308.

\bibitem{Golterman:1999au}
M.~F.~L.~Golterman and S.~Peris,
Phys. Rev. D \textbf{61} (2000), 034018.

\bibitem{Shekhovtsova:2012ra}
O.~Shekhovtsova, T.~Przedzinski, P.~Roig and Z.~Was,
Phys. Rev. D \textbf{86} (2012), 113008.

\bibitem{Antropov:2019ald}
S.~Antropov, S.~Banerjee, Z.~Was and J.~Zaremba,
[arXiv:1912.11376 [hep-ph]].

\bibitem{Miranda:2020wdg}
J.~A.~Miranda and P.~Roig,
Phys. Rev. D \textbf{102} (2020), 114017.

\bibitem{Mateu:2007tr}
V.~Mateu and J.~Portolés,
Eur. Phys. J. C \textbf{52} (2007), 325-338.

\bibitem{Weinberg:1967kj}
S.~Weinberg,
Phys. Rev. Lett. \textbf{18} (1967), 507-509

\bibitem{Kou:2018nap}
E.~Kou \textit{et al.} [Belle-II],
PTEP \textbf{2019} (2019) no.12, 123C01; PTEP \textbf{2020} (2020) 2, 029201 (erratum).


 \end{thebibliography}
\end{document}